\documentclass[sigconf,natbib=true]{acmart}
\usepackage{dsfont}
\usepackage{bbm}
\usepackage{array}
\usepackage{arydshln}
\usepackage{enumitem}
\usepackage{hyperref}
\hypersetup{
    colorlinks=true,
    linkcolor=blue,
    filecolor=magenta,      
    urlcolor=cyan,
    }

\usepackage{booktabs,multirow}
\usepackage{algorithmic}
\usepackage[linesnumbered,algoruled,boxed,lined]{algorithm2e}

\newcolumntype{P}[1]{>{\centering\arraybackslash}p{#1}}

\newcommand{\ie}{{\it i.e.}}
\newcommand{\eg}{{\it e.g.}}
\newcommand{\topN}{{top-\emph{N}}}

\newtheorem*{assumption*}{\assumptionnumber}
\providecommand{\assumptionnumber}{}


\AtBeginDocument{%
  \providecommand\BibTeX{{%
    \normalfont B\kern-0.5em{\scshape i\kern-0.25em b}\kern-0.8em\TeX}}}

\copyrightyear{2022}
\acmYear{2022}
\acmConference[SIGIR '22]{Proceedings of the 45th International ACM SIGIR Conference on Research and Development in Information Retrieval}{July 11--15, 2022}{Madrid, Spain}
\acmBooktitle{Proceedings of the 45th International ACM SIGIR Conference on Research and Development in Information Retrieval (SIGIR '22), July 11--15, 2022, Madrid, Spain}

\begin{document}
\fancyhead{}

\newcommand{\joonseok}[1]{\textcolor{blue}{\emph{[JS: #1]}}}

\title{Bilateral Self-unbiased Learning from Biased Implicit Feedback}

\author{Jae-woong Lee}
\affiliation{
  \institution{Dept. of Electrical and Computer Engineering}
  \institution{Sungkyunkwan University, Republic of Korea}
  }
\email{jwlee.icc@skku.edu}

\author{Seongmin Park}
\affiliation{
  \institution{Dept. of Artificial Intelligence}
  \institution{Sungkyunkwan University, Republic of Korea}
  }
\email{psm1206@skku.edu}
\vspace{-2mm}
\author{Joonseok Lee}
\affiliation{
  \institution{Google Research, USA}
  \institution{Seoul National University, Republic of Korea}
  }
\email{joonseok2010@gmail.com}

\author{Jongwuk Lee}\authornote{Corresponding author}
\affiliation{
  \institution{Dept. of Artificial Intelligence}
  \institution{Sungkyunkwan University, Republic of Korea}
  }
\email{jongwuklee@skku.edu}
\begin{abstract}

Implicit feedback has been widely used to build commercial recommender systems. Because observed feedback represents users' click logs, there is a semantic gap between true relevance and observed feedback. More importantly, observed feedback is usually biased towards popular items, thereby overestimating the actual relevance of popular items. Although existing studies have developed unbiased learning methods using \emph{inverse propensity weighting (IPW)} or \emph{causal reasoning}, they solely focus on eliminating the popularity bias of items. In this paper, we propose a novel unbiased recommender learning model, namely \emph{BIlateral SElf-unbiased Recommender (BISER)}, to eliminate the \emph{exposure bias} of items caused by recommender models. Specifically, BISER consists of two key components: (i) \emph{self-inverse propensity weighting (SIPW)} to gradually mitigate the bias of items without incurring high computational costs; and (ii) \emph{bilateral unbiased learning (BU)} to bridge the gap between two complementary models in model predictions, \ie, user- and item-based autoencoders, alleviating the high variance of SIPW. Extensive experiments show that BISER consistently outperforms state-of-the-art unbiased recommender models over several datasets, including Coat, Yahoo! R3, MovieLens, and CiteULike.

\end{abstract}





\maketitle

\section{Introduction}\label{sec:Introduction}

Collaborative filtering (CF)~\cite{RicciRS15,ChoiJLL21,LeeKLS13,AdomaviciusT05} is the most prevalent technique for building commercial recommender systems. CF typically utilizes two types of user feedback: \emph{explicit} and \emph{implicit feedback}. Explicit feedback provides richer information about user preferences than implicit feedback as users explicitly rate how much they like or dislike the items. However, it is difficult to collect explicit feedback from various real-world applications because only a few users provide feedback after experiencing the items. On the other hand, implicit feedback is easily collected by recording various users' behaviors, \eg, clicking a link, purchasing a product, or browsing a web page.

There are several challenges in using implicit feedback in CF. (i) Existing studies~\cite{HuKV08,HeLZNHC17,XueDZHC17,HeDWTTC18,TangW18a,WuDZE16,YaoTYXZSL19,ShenbinATMN20,PrathamaSYW21,ChoiKLSL22} regard the observed user interactions solely as positive feedback. However, some observed interactions, such as clicking an item or viewing a page, do not necessarily indicate whether the user likes the item; there is a semantic gap between the true relevance and the observed interactions. (ii) User feedback is observed at uniformly not random. For instance, users tend to interact more with popular items, so the more popular the items are, the more they are collected in the training dataset. Because of the inherent nature of implicit user-item interactions, recommender models are biased toward ranking popular items with high priority.


Existing studies have developed unbiased recommender learning methods~\cite{SaitoYNSN20,Saito20,ZhuHZC20,QinCMNQW20} to estimate \emph{true} user preferences from implicit feedback under the \emph{missing-not-at-random (MNAR)} assumption~\cite{MarlinZRS07,Steck10,ZhengGLHLJ21}. They formulated a new loss function to eliminate the bias of items by using \emph{inverse propensity weighting (IPW)}, which has been widely established in causal inference~\cite{JoachimsSS17,WangGBMN18,AiBLGC18,HuWPL19,Wang0WW21,WangBMN16,OvaisiAZVZ20,SchnabelSSCJ16,0003ZS021,WangZSQ19,WangZSQ19,SaitoYNSN20,Saito20,ZhuHZC20,QinCMNQW20}. In addition, recent studies~\cite{ZhangFHWSLZ21, WeiFCWYH21} introduced a \emph{causal graph} that represents a cause effect relationship for recommendations and removes the effect of item popularity.

Specifically, they are categorized in two directions. The first approach exploits a heuristic function to eliminate the popularity bias of items. Although item popularity is a critical factor in the bias of training datasets, there are other vital factors, such as exposure bias, in the recommended models. To overcome this issue, the second approach develops a learning-based method that accounts for various bias factors. Joint learning methods~\cite{Saito20at, ZhuHZC20} first suggested utilizing a pseudo-label or propensity score inferred from an additional model. \citet{ZhuHZC20} employed multiple models for different subsets of a training dataset to infer the propensity scores. \citet{Saito20at} adopted two pre-trained models with other parameter initializations and generated a pseudo-label as the difference between the two model predictions. They then made use of consistent predictions by training multiple models. However, as multiple models converge to a similar output, it leads to an estimation overlap issue. Recently, causal graph-based training methods~\cite{ZhangFHWSLZ21, WeiFCWYH21} were proposed to overcome the sensitivity of IPW strategies. \citet{WeiFCWYH21} modeled a causal graph using item popularity and user conformity to predict true relevance. \citet{ZhangFHWSLZ21} analyzed the negative effects of item popularity through a causal graph and removed bias through causal intervention. However, they did not address the exposure bias caused by recommender models.


To eliminate exposure bias, we propose a novel unbiased recommender learning model, namely the \emph{BIlateral SElf-unbiased Recommender (BISER)} with two key components: \emph{self-inverse propensity weighting (SIPW)} and \emph{bilateral unbiased learning (BU)}. Motivated by self-distillation~\cite{MobahiFB20, YangXSY19}, we first devise a \emph{self-inverse propensity weighting (SIPW)} to iteratively mitigate the exposure bias of items. Specifically, we reuse the model prediction from the previous training iteration, enabling us to gradually eliminate the exposure bias of items as the training evolves. Notably, SIPW has two key advantages: (i) it effectively handles the exposure bias caused by recommender models and (ii) it does not require an additional model for propensity score estimation.

IPW usually suffers from the high variance problem, as reported in the literature~\cite{ZhangFHWSLZ21, WeiFCWYH21}. To resolve this issue, we first assume that the true user preference should be consistent with the predictions of different models. We then design \emph{bilateral unbiased learning (BU)} using two recommender models. Specifically, we utilize user- and item-based autoencoders~\cite{SedhainMSX15}. Because they capture different hidden patterns on the user and item sides, it does not require us to split the training and estimation subsets from an entire training set~\cite{ZhuHZC20} or to utilize multiple models with different parameter initializations~\cite{Saito20at}. We exploit the predicted value of one model as a pseudo-label for another model. As a result, we resolve the high-variance issue in estimating the SIPW.

To summarize, the key contributions of this paper are as follows:
\begin{itemize}[leftmargin=5mm]
    \item We review existing studies on unbiased recommender learning and analyze their limitations to eliminate the exposure bias of items caused by recommender models (Section~\ref{sec:background}).
    
    \vspace{0.5mm}
    \item We propose a novel unbiased recommender learning model, namely \emph{BIlateral SElf-unbiased Recommender (BISER)}, utilizing (i) a learning-based propensity score estimation method, \ie, \emph{self-inverse propensity weighting (SIPW)}, and (ii) \emph{bilateral unbiased learning (BU)} using user- and item-based autoencoders with complementary relationships (Section~\ref{sec:model}).
    

    \vspace{0.5mm}
    \item We demonstrate that the BISER outperforms state-of-the-art unbiased recommender models, including RelMF~\cite{SaitoYNSN20}, AT~\cite{Saito20at}, CJMF~\cite{ZhuHZC20}, PD~\cite{ZhangFHWSLZ21}, and MACR~\cite{WeiFCWYH21}, on both unbiased evaluation datasets (\eg, Coat and Yahoo! R3) and conventional datasets (\eg, MovieLens and CiteULike) (Sections~\ref{sec:setup}--\ref{sec:result}).
\end{itemize}
\begin{figure*}
\includegraphics[height=3.4cm]{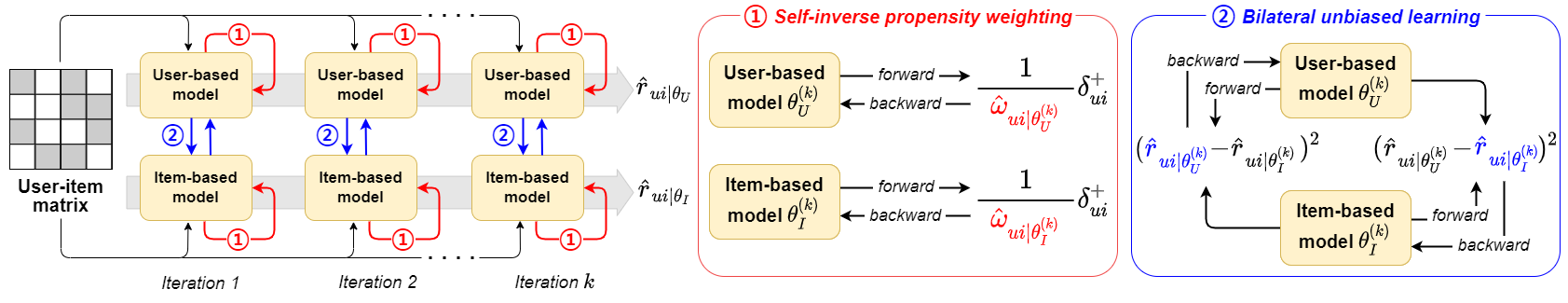} \\ 
\vspace{-1mm}
\caption{Overall training process of BISER. It consists of two parts, self-inverse propensity weighting (SIPW) and bilateral unbiased learning (BU), where SIPW is computed by using recommender model's own output, and BU is used to reduce the gap between the predictions of user- and item-based models.}\label{fig:proposed}

\vspace{-3mm}
\end{figure*}
\begin{figure}[t]
\centering
\begin{tabular}{cc}
\includegraphics[width=0.17\textwidth]{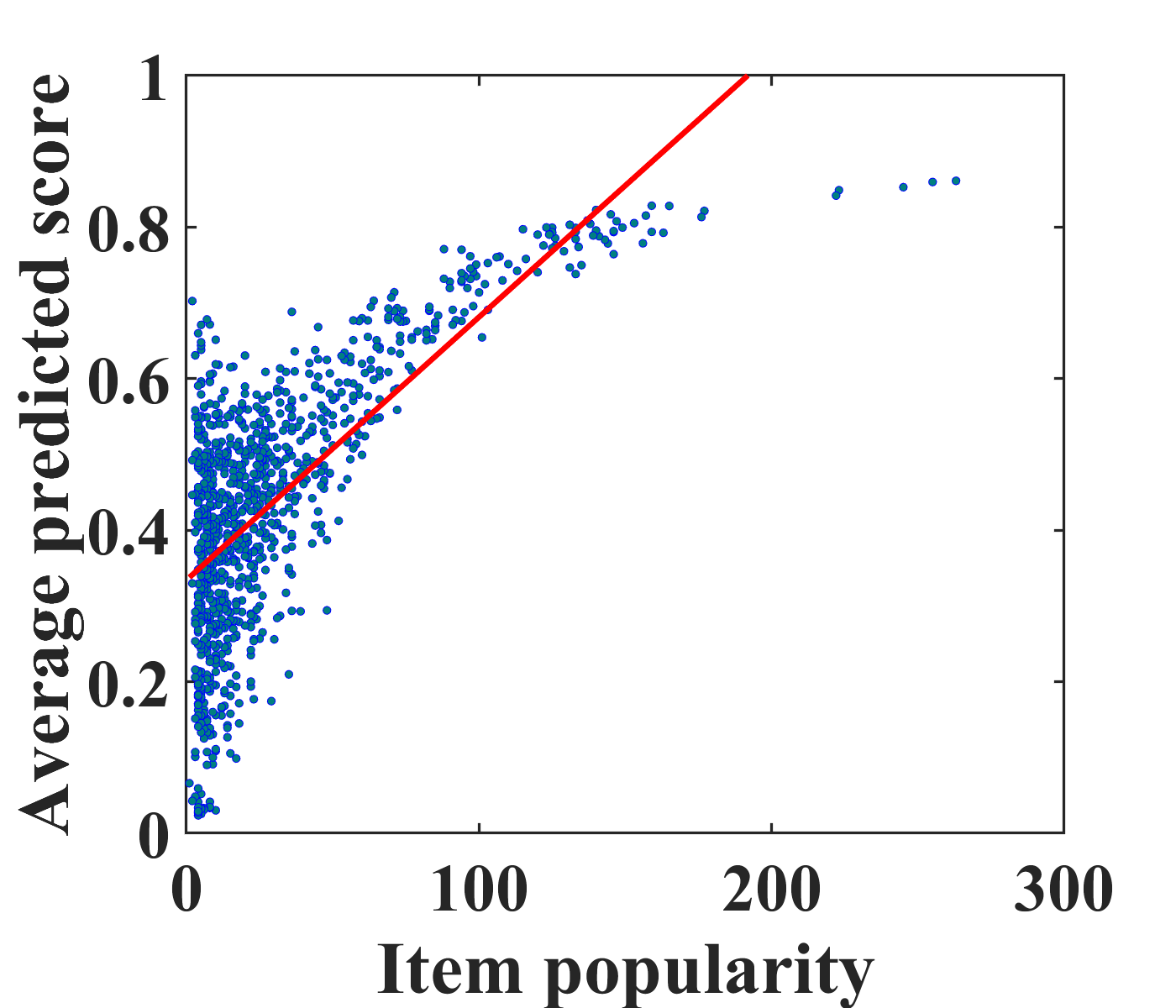}\quad &\quad
\includegraphics[width=0.17\textwidth]{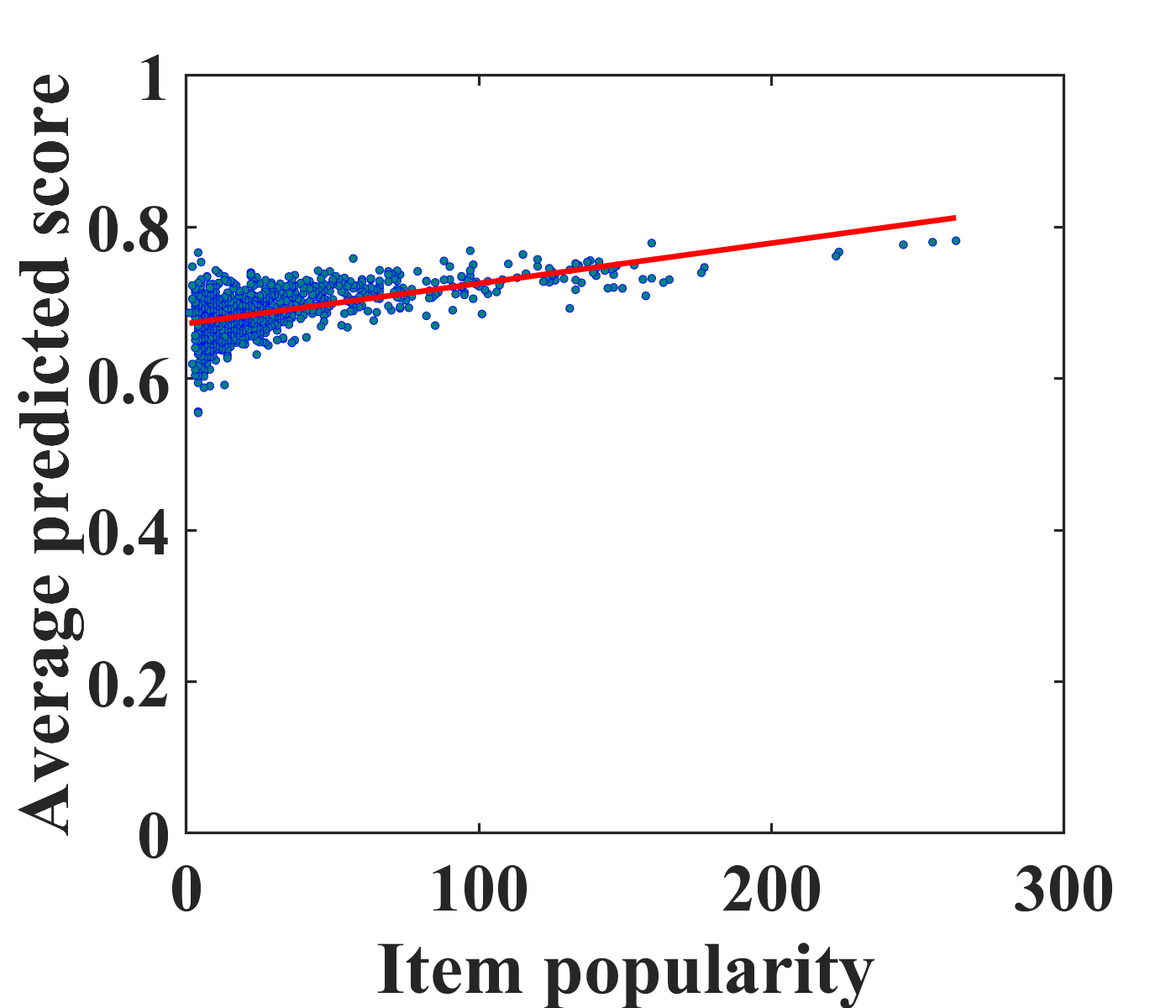} \\
(a) MF \quad & \quad (b) Unbiased MF (ours) \\

\includegraphics[width=0.17\textwidth]{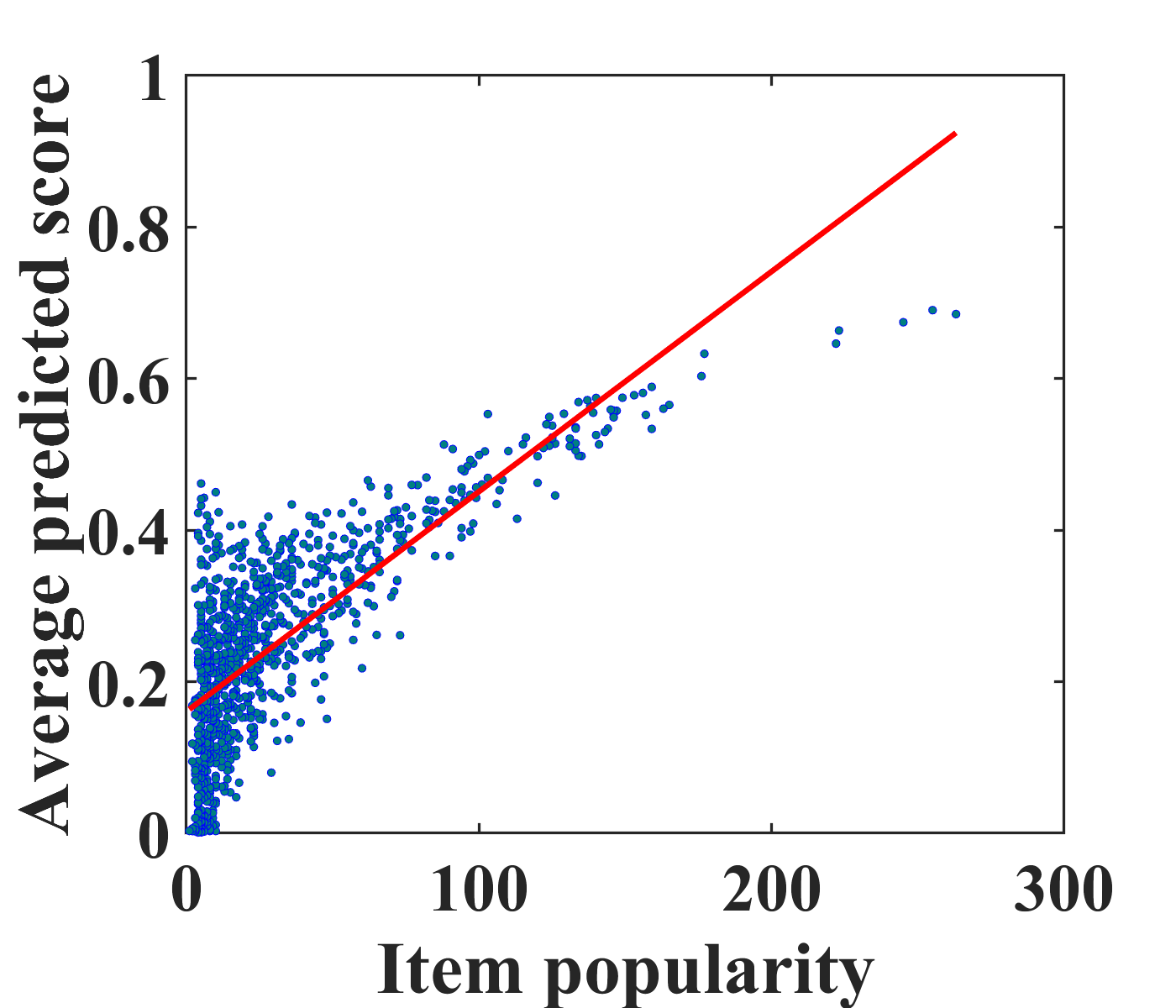}  \quad&\quad
\includegraphics[width=0.17\textwidth]{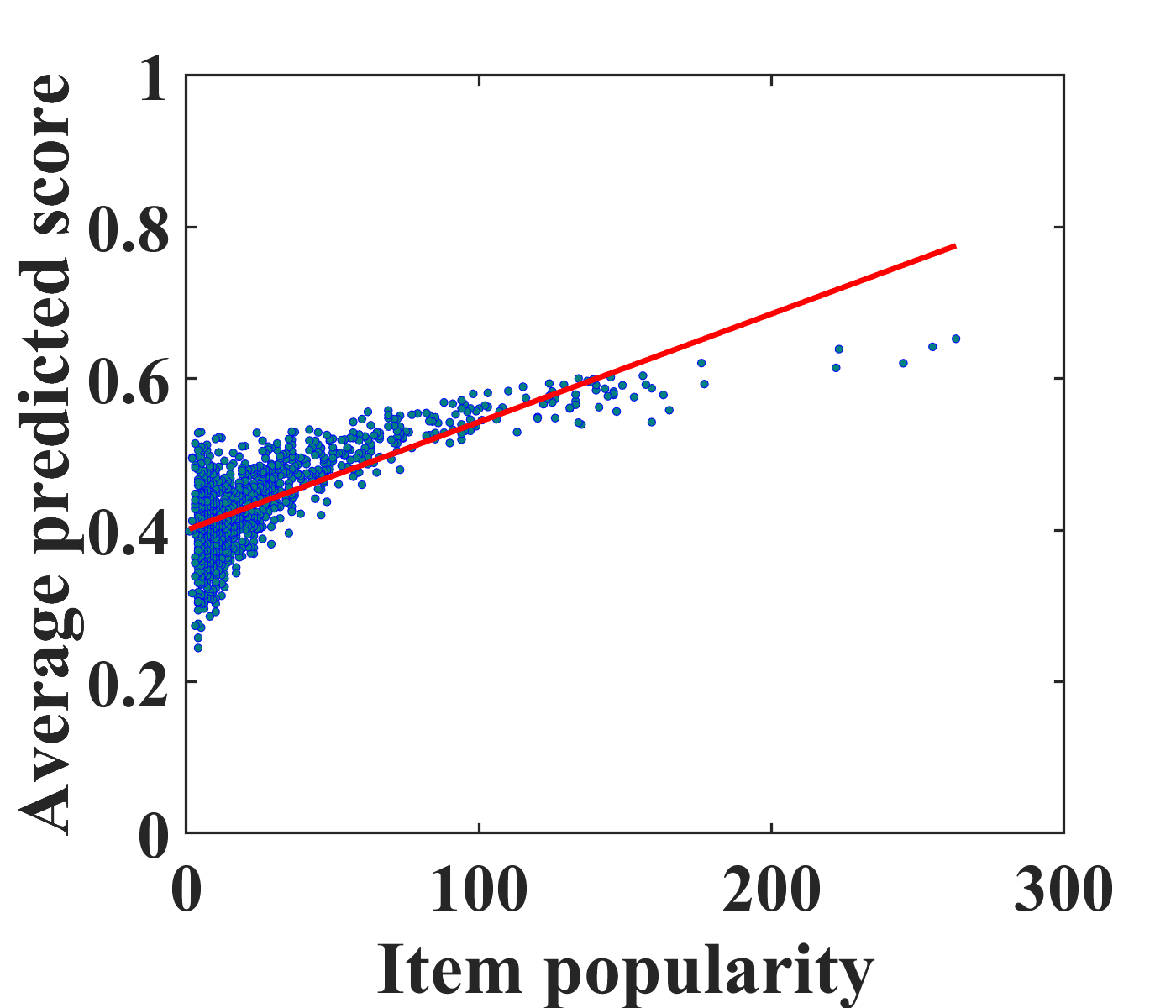} \\
(c) AE \quad & \quad (d) Unbiased AE (ours) \\
\end{tabular}
\vspace{-1mm}
\caption{Correlation between the number of ratings, \ie, popularity, and the average model prediction score for each item on the ML-100K dataset. While existing models, \ie, MF and AE, tend to predict high scores for popular items without addressing the bias of items, our proposed model can mitigate the bias of items, showing that the distribution for predicted scores of items is less skewed towards popularity.}\label{fig:motive_fig}
\vspace{-4mm}
\end{figure}

\section{Background}\label{sec:background}

\textbf{Notations}. Formally, we denote $\mathcal{U}$ as a set of $m$ users and $\mathcal{I}$ as a set of $n$ items. We are given a user-item \emph{click matrix} $\textbf{Y} \in \{0,1\}^{m \times n}$, \begin{equation}
y_{ui} = \left\{
  \begin{array}{ll}
    1 & \text{if user $u$ interacted with item $i$};\\
    0 & \text{otherwise}.
  \end{array}
\right.
\end{equation}
We model $y_{ui}$ as a Bernoulli random variable, indicating the interaction of a user $u$ on an item $i$.
For implicit feedback data, the user interaction is a result of observation and preference. That is, a user may click on an item if (i) the item is exposed to the user and the user is aware of the item (\emph{exposure}) and (ii) the user is actually interested in the item (\emph{relevance}). Existing studies~\cite{SchnabelSSCJ16,Saito20,SaitoYNSN20,ChenDWFWH20,ZhuHZC20,QinCMNQW20} formulate this idea as follows:
\begin{equation}\label{eq:click}
P(y_{ui} = 1) = P(o_{ui} = 1) \cdot P(r_{ui} = 1)  = \omega_{ui} \cdot \rho_{ui},
\end{equation}
where $o_{ui}$ is an element of the \emph{observation matrix} $\textbf{O} \in \{0, 1\}^{m \times n}$ that represents whether the user $u$ has observed the item $i$ ($o_{ui} = 1$) or not ($o_{ui} = 0$), and $r_{ui}$ is an element of the \emph{relevance matrix} $\textbf{R} \in \{0, 1\}^{m \times n}$, representing true relevance regardless of observance. If the user $u$ likes the item $i$, $r_{ui} = 1$, and $r_{ui} = 0$ otherwise. For simplicity, we denote $P(o_{ui} = 1)$ and $P(r_{ui} = 1)$ with $\omega_{ui}$ and $\rho_{ui}$, respectively. The interaction matrix $\textbf{Y}$ with biased user behavior is decomposed into element-wise multiplication of the observation and relevance components. 

\vspace{1mm}
\noindent
\textbf{Unbiased recommender learning}. Our goal is to learn an unbiased ranking function from implicit feedback under the MNAR assumption. Although there are various loss functions for training recommender models, such as \emph{point-wise}, \emph{pair-wise}, and \emph{list-wise} losses~\cite{Rendle21}, we use the point-wise loss function in this paper. Given a set of user-item pairs $\mathcal{D} = \mathcal{U} \times \mathcal{I}$, the loss function for biased interaction data is
\begin{equation}
\label{eq:click_loss}
  \mathcal{L}_\text{biased}(\hat{\textbf{R}}) = \frac{1}{|\mathcal{D}|}\sum_{(u, i) \in \mathcal{D}}
  \left( y_{ui} \delta^{+}_{ui} + \left( 1 - y_{ui} \right) \delta^{-}_{ui} \right),
\end{equation}
where $\hat{\textbf{R}}$ is the prediction matrix for $\textbf{R}$, and $\delta^{+}_{ui}$ and $\delta^{-}_{ui}$ are the loss of user $u$ on item $i$ for positive and negative preference, respectively. Under the point-wise loss setting, we adopt either the cross-entropy or the sum-of-squared loss. With the cross-entropy, for example, $\delta^{+}_{ui} = -\log{(\hat{r}_{ui})} $ and $\delta^{-}_{ui}= -\log{(1 - \hat{r}_{ui})}$.

Similarly, the ideal loss function that relies purely on relevance is formulated as follows:
\begin{equation}
\label{eq:ideal_loss}
  \mathcal{L}_\text{ideal}(\hat{\textbf{R}}) = \frac{1}{|\mathcal{D}|}\sum_{(u, i) \in \mathcal{D}}
  \left( \rho_{ui} \delta^{+}_{ui} + \left( 1 - \rho_{ui} \right) \delta^{-}_{ui} \right),
\end{equation}
where the biased observation $y_{ui}$ is substituted with the pure relevance $\rho_{ui} = P(r_{ui} = 1)$.

Because user interaction data are typically sparse and collected under the MNAR assumption, it is necessary to bridge the gap between the loss functions for clicks and relevance. \citet{SaitoYNSN20} proposed a loss function using \emph{inverse propensity weighting (IPW)}:
\begin{equation}
\label{eq:saito}
  \mathcal{L}_\text{unbiased}(\hat{\textbf{R}}) = \frac{1}{|\mathcal{D}|}\sum_{(u, i) \in \mathcal{D}}{ \left(\frac{y_{ui}}{\omega_{ui}} \delta^{+}_{ui} + \left ( 1-\frac{y_{ui}}{\omega_{ui}} \right )\delta^{-}_{ui} \right)},
\end{equation}
where $\omega_{ui}$ is the \emph{inverse propensity score} that indicates the probability of observing the item $i$ by the user $u$. As proved in~\cite{SaitoYNSN20}, the expectation of the unbiased loss function in Eq.~\eqref{eq:saito} is equivalent to the ideal loss function in Eq.~\eqref{eq:ideal_loss}:
\begin{align}
  \mathbb{E}\left [ \mathcal{L}_\text{unbiased}(\hat{\textbf{R}}) \right ] &= \frac{1}{|\mathcal{D}|}\sum_{(u, i) \in \mathcal{D}}{ \left (\frac{\mathbb{E}\left [ y_{ui} \right ]}{\omega_{ui}} \delta^{+}_{ui} + \left ( 1-\frac{\mathbb{E}\left [ y_{ui} \right ]}{\omega_{ui}} \right )\delta^{-}_{ui} \right )} \nonumber \\
  & = \frac{1}{|\mathcal{D}|}\sum_{(u, i) \in \mathcal{D}}{ \left (\rho_{ui} \delta^{+}_{ui} + \left ( 1-\rho_{ui} \right ) \delta^{-}_{ui} \right )}.
\end{align}

The critical issue is how to estimate the propensity score $\omega_{ui}$ from observed feedback (\eg, clicks). Previous studies~\cite{SaitoYNSN20, Saito20, ZhuHZC20, QinCMNQW20, LeePL21} have developed several solutions to estimate $\omega_{ui}$. First, \cite{SaitoYNSN20, Saito20, LeePL21} introduced a heuristic function to estimate item popularity without using an additional model. Although intuitive, it focuses only on addressing item popularity and is thus, incapable of handling exposure bias caused by recommender models. Second, additional models were used to estimate the propensity score~\cite{ZhuHZC20, QinCMNQW20}. However, this incurs high computational costs, as they mostly utilize three or more models.
\section{BISER: Proposed Model}
\label{sec:model}

To motivate unbiased recommender modeling, we first investigate the bias from conventional recommendation models, \ie, matrix factorization (MF) and autoencoders (AE)~\cite{SedhainMSX15} on the MovieLens-100K (ML-100K) dataset. Figures~\ref{fig:motive_fig}(a) and (c) depict a positive correlation between the average predicted scores for clicked items and the number of ratings in the training set, \ie, popularity. From this case study, we confirm that popular items tend to be recommended to more users with conventional models. Even worse, the biased recommendations can exacerbate the bias of the training dataset as user feedback is continually collected.


To address this problem, it is vital to eliminate the exposure bias caused by recommender models. Existing studies~\cite{Saito20,SaitoYNSN20} have mainly focused on modeling the popularity bias of items. Because user experiences are mostly biased towards the recommended items, we aim to eliminate exposure bias caused by recommender models. We estimate the propensity score of the items and eliminate exposure bias during model training. Figures~\ref{fig:motive_fig}(b) and (d) show the effectiveness of the proposed unbiased learning method. It is clearly observed that the correlation between the actual and predicted ratings in the proposed model is weaker than that in the traditional models. This pilot study indicates that our unbiased model effectively reduces exposure bias caused by recommender models.

\subsection{Model Architecture}
\label{sec:overview}

In this section, we present the novel unbiased recommender model, namely, the \emph{Bilateral Self-unbiased Recommender (BISER)}. Specifically, it consists of two parts: \emph{self-inverse propensity weighting (SIPW)} and \emph{bilateral unbiased learning (BU)}, as shown in Figure~\ref{fig:proposed}. Motivated by self-distillation~\cite{MobahiFB20,YangXSY19}, which extracts knowledge of model predictions during model training, we estimate inverse propensity weighting (IPW) in an iterative manner. Thus, the proposed model gradually eliminates the exposure bias of items during model training. It is efficient because it neither requires a pre-defined propensity scoring function nor an additional model for propensity estimation.

In this process, we adopt two recommender models with complementary characteristics, user- and item-based autoencoders that capture heterogeneous semantics and relational information from the users' and items' perspectives, respectively. This bypasses the estimation overlap issue that occurs when the outputs of multiple models with the same structure and trained on the same training sets, converge similarly. Furthermore, we emphasize that our proposed approach is model-agnostic; two or more models may be used as long as they convey heterogeneous signals from the user-item interaction data.

\vspace{1mm}
\noindent
\textbf{Self-inverse propensity weighting (SIPW).} Existing studies deal with eliminating the exposure bias in two ways: (i) by modeling only item popularity, which affects the clicking behavior of users, regardless of an additional estimator~\cite{SaitoYNSN20,Saito20}, or (ii) by adopting an additional model to estimate the exposure bias by interleaving relevance and observation estimation~\cite{ZhuHZC20}. \citet{ZhuHZC20} split an entire dataset into multiple training and estimation subsets, incurring high computational costs.

In this paper, we formulate a new bias estimation method, namely \emph{self-inverse propensity weighting (SIPW)}. First, we introduce an ideal unbiased recommender model using interaction data. Given a recommender list of items $\pi_{u}$ to user $u$, the probability of the user interacting with an item $i$ is given by
\begin{equation}
  P^*(y_{ui} = 1 | \pi_{u}) = P^*(o_{ui} = 1 | \pi_{u}) \cdot P(r_{ui} = 1 | \pi_{u}),
\end{equation}
where $P^*(o_{ui} = 1 | \pi_{u})$ is the ideal probability of observing an item $i$, \ie, \emph{completely-at-random distribution}. Besides, we estimate the probability that the user $u$ interacts with the item $i$ by the biased recommender model as
\begin{equation}
  \hat{P}(y_{ui} = 1 | \pi_{u}) = \hat{P}(o_{ui} = 1 | \pi_{u}) \cdot P(r_{ui} = 1 | \pi_{u}),
\end{equation}
where $\hat{P}(o_{ui} = 1 | \pi_{u})$ is the estimated probability of observing an item $i$.

By combining them, we estimate the IPS as the ratio between the two probability distributions for interactions. Note that this formulation is closely related to the unbiased learning-to-rank method~\cite{WangBMN16}.
\begin{equation}
\frac{P^{*}(y_{ui} = 1 | \pi_{u})}{\hat{P}(y_{ui} = 1 | \pi_{u})} = \frac{P^*(o_{ui} = 1 | \pi_{u}) \cdot P(r_{ui} = 1 | \pi_{u})}{\hat{P}(o_{ui} = 1 | \pi_{u}) \cdot P(r_{ui} = 1 | \pi_{u})}
\end{equation}

Assuming $P^*(o_{ui} = 1 | \pi_{u})$ follows uniform distribution, the probability for every item in $\pi_{u}$ is simply regarded as a constant. Therefore, we estimate a propensity score $\hat{\omega}_{ui}$ as
\begin{equation}
  \frac{P^{*}(y_{ui} = 1 | \pi_{u})}{\hat{P}(y_{ui} = 1 | \pi_{u})} \propto \frac{1}{\hat{P}(o_{ui} = 1 | \pi_{u})} = \frac{1}{\hat{\omega}_{ui}}.
  \label{eq:click_proposed}
\end{equation}



The remaining issue is how to estimate the probability $\hat{\omega}_{ui}$ of observing item $i$ by user $u$. Inspired by self-distillation~\cite{MobahiFB20,YangXSY19} that utilizes the knowledge of model predictions during model training, we reuse the model prediction as $\hat{\omega}_{ui}$. Despite its simplicity, self-propensity estimation has several benefits. (i) Our SIPW method stably estimates $\hat{\omega}_{ui}$ by regularizing prior knowledge, gradually removing the bias. (ii) It does not require an additional inference process to estimate the propensity scores. (iii) Owing to the model-agnostic property, our solution can be applied with various recommender models, \eg, MF~\cite{HuKV08} and AE~\cite{SedhainMSX15}.

Formally, we formulate an unbiased loss function using SIPW as
\begin{equation}\label{eq:self_unbiased_epoch}
\begin{aligned}
\mathcal{L}_{SIPW}(& \hat{\textbf{R}}; \theta^{(k)}) = \\    & \frac{1}{|\mathcal{D}|}\sum_{(u, i) \in \mathcal{D}}{ \left(\frac{y_{ui}}{\hat{\omega}_{ui | \theta^{(k)}}} \delta^{+}_{ui} + \left ( 1-\frac{y_{ui}}{\hat{\omega}_{ui | \theta^{(k)}}} \right )\delta^{-}_{ui} \right)},
\end{aligned}
\end{equation}
where $\theta^{(k)}$ is the model parameter at the $k$-th iteration and $\hat{\omega}_{ui | \theta^{(k)}}$ is the self-inverse propensity score at the $k$-th iteration.

\vspace{1mm}
\noindent
\textbf{Bilateral unbiased learning (BU).} Although IPW is theoretically principled, it often leads to suboptimal performance owing to the high variance problem in practice~\cite{SwaminathanJ15,GilotteCNAD18,WangZSQ19}. An existing study~\cite{Saito20at} trains multiple models where the prediction from one model is used as the pseudo-label for the other models. The pseudo-label is then used to regularize the original model. Although it helps mitigate the high-variance problem, multiple models can show a similar tendency as they are trained.

To resolve this issue, we utilize two recommender models with different characteristics. Unifying user- and item-based recommender models improves predictions in conventional recommender models~\cite{WangVR06,YamashitaKS11,ZhuWC19}. As they tend to capture different aspects of user and item patterns, adopting heterogeneous models helps to relax the estimation overlap issue.

Specifically, we utilize user- and item-based autoencoders to discover unique and compensating patterns from complex user-item interactions. Inspired by label shift~\cite{LiptonWS18,Azizzadenesheli19}, the two models should converge to the same value to correctly estimate the true relevance.
\begin{equation}
  \label{eq:assumption2}
  P(r_{ui} = 1 | y_{ui} = 1, \theta_U) = P(r_{ui} = 1 | y_{ui} = 1, \theta_I),
\end{equation}
where $\theta_U$ and $\theta_I$ are the parameters of user- and item-based autoencoders, respectively~\cite{SedhainMSX15}. Finally, we present a loss function for bilateral unbiased learning using the two model predictions.
\begin{equation}
  \label{eq:co_unbiased}
  \mathcal{L}_{BU}(\hat{\textbf{R}}; \theta_U^{(k)}, \theta_I^{(k)}) = \frac{1}{|\tilde{\mathcal{D}}|}\sum_{(u, i) \in \tilde{\mathcal{D}}}{ \left( \hat{r}_{ui | \theta_U^{(k)}} - \hat{r}_{ui | \theta_I^{(k)}}\right)^2},
\end{equation}
where $\hat{r}_{ui | \theta_U^{(k)}}$ and $\hat{r}_{ui | \theta_I^{(k)}}$ are predictions from each model at the $k$-th iteration, respectively. $\tilde{\mathcal{D}}$ is a set of observed user-item pairs.


\subsection{Training and Inference}
\label{sec:learning}
\noindent
Given user- and item-based autoencoders, each model is trained from scratch using SIPW. At each iteration, we update the model parameters by minimizing the loss in Eq.~\eqref{eq:self_unbiased_epoch}. We then account for the loss function to minimize the difference between the two model predictions. Specifically, the user-based model with parameter $\theta_U^{(k)}$ is trained with the predictions from the item-based model with parameter $\theta_I^{(k)}$ as pseudo-labels, and vice versa. Therefore, we can reduce the high-variance issue of SIPW by using two model predictions.

Based on Eq.~\eqref{eq:self_unbiased_epoch} and~\eqref{eq:co_unbiased}, we represent the final loss function to simultaneously train two models at the $k$-th iteration.
\begin{equation}
\label{eq:proposed_final_1}
  \mathcal{L}_{UAE}(\hat{\textbf{R}}; \theta_U^{(k)}) = \mathcal{L}_{SIPW}(\hat{\textbf{R}}; \theta_U^{(k)}) + \lambda_U\mathcal{L}_{BU}(\hat{\textbf{R}}; \theta_U^{(k)}, \theta_I^{(k)}),
\end{equation}
\begin{equation}
\label{eq:proposed_final_2}
  \mathcal{L}_{IAE}(\hat{\textbf{R}}; \theta_I^{(k)}) = \mathcal{L}_{SIPW}(\hat{\textbf{R}}; \theta_I^{(k)}) + \lambda_I\mathcal{L}_{BU}(\hat{\textbf{R}}; \theta_I^{(k)}, \theta_U^{(k)}),
\end{equation}

\noindent
where $\lambda_U$ and $\lambda_I$ are the hyperparameters to control the importance of $\mathcal{L}_{BU}$. 



Once model training is terminated, the biases of the two models are eliminated from the users' and items' perspectives. By unifying user preferences from different perspectives, we can improve model predictions~\cite{WangVR06,YamashitaKS11,ZhuWC19}. Finally, we use the average of the predictions of the two models for the final rating.
\begin{equation}
  \label{eq:inferofBISEL}
  \hat{r}_{ui} = \frac{\hat{r}_{ui | \theta_U} + \hat{r}_{ui | \theta_I}}{2}.
\end{equation}

\SetKwInOut{Parameter}{Parameters}
\vspace{3mm}

\begin{algorithm}[t]
  \KwIn{Dataset $\mathcal{D} = \{ \tilde{\mathcal{D}}, \mathcal{D}\backslash\tilde{\mathcal{D}} \}$, hyper-parameters $\lambda_U$, $\lambda_I$, the number of iterations $K$}
  
  \KwOut{Predicted matrix $\hat{\textbf{R}}$}

  Initialize $\theta_U^{(0)}$ and $\theta_I^{(0)}$ for user- and item-based AE.
  \label{alg:algorithm1}

      
  \For{$k=1$ to $K$}
  {
      Update $\theta_U^{(k)}$ for the user-based AE using Eq.~\eqref{eq:proposed_final_1}.
      
      Update $\theta_I^{(k)}$ for the item-based AE using Eq.~\eqref{eq:proposed_final_2}.

  }
  Compute $\hat{\textbf{R}}$ for user- and item-based AE using Eq.~\eqref{eq:inferofBISEL}.
  
  \Return{$\hat{\textbf{R}}$}
  
  \caption{Bilateral self-unbiased recommender (BISER)}
\end{algorithm}

The pseudo-code of BISER with user- and item-based AE is described in Algorithm~\ref{alg:algorithm1}. In the learning process, it first initializes the parameters $\theta_U$ and $\theta_I$ of both AE (line 1). Subsequently, $\theta_U$ and $\theta_I$ are updated using Eqs.~\eqref{eq:proposed_final_1} and ~\eqref{eq:proposed_final_2} with the model predictions from the previous iteration, respectively (lines 2-5). At each iteration, it calculates the predictions of both models for all the clicked user-item pairs. After the two-model training is terminated, we obtain the predicted matrix $\hat{\textbf{R}}$ using the two model predictions (lines 6 and 7).

\section{Experimental Setup}
\label{sec:setup}

\noindent
\textbf{Datasets and preprocessing}. Table~\ref{tab:statistics} summarizes the statistics of the datasets used in the evaluation. Among these datasets, the Coat and Yahoo! R3 datasets are specially designed to evaluate an unbiased setting. Specifically, their training set collected user feedback without special treatment (so it is likely to be biased), but the test set was carefully designed to be unbiased by explicitly asking for feedback for a pre-selected random set of items from each user. These two datasets are ideal for evaluating unbiased recommender models in a controlled setting. We also used four benchmark datasets, \ie, MovieLens (ML)-100K, ML-1M, ML-10M, and CiteULike, where the training and test datasets are inherently collected with click bias.

To account for implicit feedback, five datasets with explicit feedback (\ie, Coat\footnote{\url{https://www.cs.cornell.edu/~schnabts/mnar/}}, Yahoo! R3\footnote{\url{http://webscope.sandbox.yahoo.com/}}, MovieLens\footnote{\url{http://grouplens.org/datasets/movielens/}} (ML)-100K, ML-1M, and ML-10M) are converted into implicit feedback. We treat the items with four or higher scores as positive feedback and the remaining ratings are regarded as missing feedback. Unlike the other five datasets, CiteULike~\cite{WangWY15} provides implicit feedback.

We conduct the following experiments with different settings:
\begin{itemize}[leftmargin=5mm]
  \item \textbf{MNAR-MAR}: Using Coat and Yahoo! R3, we train the model on the training dataset collected under the MNAR assumption and evaluate it on the test set collected under MAR. The training and test sets are provided separately. Within the training set, we set 30\% of the ratings per user for validation purposes. The test set of the Coat and Yahoo! R3 datasets consists of 16 and 10 item ratings per user, respectively. For the items in the test set, we retrieve the top-$N$ items.
  \vspace{0.5mm}

  \item \textbf{MNAR-MNAR}: Using four traditional datasets, we train and evaluate the recommender models under the MNAR assumption. Before splitting the dataset, we removed users who rated 10 or fewer items and items rated by 5 or fewer users. For each user, we randomly held 80\% and 20\% as the training and test sets, respectively, and further split 30\% of the training set into the validation set. Then, we evaluate the top-$N$ recommendation for all unrated items.
\end{itemize}

\makeatletter
\def\thickhline{%
  \noalign{\ifnum0=`}\fi\hrule \@height \thickarrayrulewidth \futurelet
   \reserved@a\@xthickhline}
\def\@xthickhline{\ifx\reserved@a\thickhline
               \vskip\doublerulesep
               \vskip-\thickarrayrulewidth
             \fi
      \ifnum0=`{\fi}}
\makeatother

\newlength{\thickarrayrulewidth}
\setlength{\thickarrayrulewidth}{1.5\arrayrulewidth}


\begin{table}[t]
\centering
\caption{Statistics of the datasets after preprocessing.}\label{tab:statistics}
\begin{center}
\renewcommand{\arraystretch}{0.9} 
\begin{tabular}{P{1.5cm}P{1.0cm}P{1.0cm}P{1.5cm}P{1.0cm}}
\toprule
\multicolumn{1}{c}{Datasets} & \multicolumn{1}{c}{\#users} & \multicolumn{1}{c}{\#items} & \multicolumn{1}{c}{\#interactions}  & \multicolumn{1}{c}{Sparsity}\\
\hline
Coat & 290 & 300 & 1,905 & 0.978 \\
Yahoo! R3 & 15,400 & 1,000 & 125,077 & 0.992 \\
\hline
ML-100K & 897 & 1,007 & 54,103 & 0.940 \\
ML-1M & 5,950 & 3,125 & 573,726 & 0.969 \\
ML-10M & 66,028 & 8,782 & 4,977,095 & 0.991 \\
CiteULike & 5,551 & 15,452 & 205,813 & 0.998 \\
\bottomrule
\end{tabular}
\end{center}
\vspace{-3mm}
\end{table}

\begin{table*}[t]
\centering
\caption{Accuracy comparison of ours and baseline models on Coat and Yahoo! R3. The best is in bold font, * and ** indicate $p < 0.005$ and $p < 0.001$ for a one-tailed t-test,  $\dagger$ indicates the best baseline model, \ie, CJMF for t-test and performance gain. Gain indicates an improvement ratio to CJMF. The results are averaged over 10 and 5 runs on Coat and Yahoo! R3, respectively.}\label{tab:unbiased}
\begin{center}
\renewcommand{\arraystretch}{0.95} 
\begin{tabular}{P{1.5cm}P{1.8cm}P{1.0cm}P{1.0cm}P{1.0cm}P{1.0cm}P{1.0cm}P{1.0cm}P{1.0cm}P{1.0cm}P{1.0cm}}
\toprule

\multicolumn{2}{c}{} & \multicolumn{3}{c}{NDCG$@N$} & \multicolumn{3}{c}{MAP$@N$} & \multicolumn{3}{c}{Recall$@N$}\\
\cmidrule(r){3-5}
\cmidrule(lr){6-8}
\cmidrule(l){9-11}
 Datasets & Models & $N$ = 1 & $N$ = 3 & $N$ = 5 & $N$ = 1 & $N$ = 3 & $N$ = 5 & $N$ = 1 & $N$ = 3 & $N$ = 5  \\
\hline
\multirow{10}{*}{Coat}
& MF~\cite{HuKV08} & 0.3748 & 0.3441 & 0.3714 & 0.1346 & 0.2100 & 0.2566 & 0.1346 & 0.2592 & 0.3705 \\
& UAE~\cite{SedhainMSX15} & 0.3610 & 0.3546 & 0.3815 & 0.1265 & 0.2165 & 0.2648 & 0.1265 & 0.2785 & 0.3869 \\
& IAE~\cite{SedhainMSX15} & 0.3655 & 0.3560 & 0.3812 & 0.1311 & 0.2185 & 0.2651 & 0.1311 & 0.2769 & 0.3847 \\
\cline{2-11}
& RelMF~\cite{SaitoYNSN20} & 0.3959 & 0.3659 & 0.3922 & 0.1484 & 0.2281 & 0.2758 & 0.1484 & 0.2819 & 0.3926 \\
& AT~\cite{Saito20at} & 0.4017 & 0.3652 & 0.3912 & 0.1517 & 0.2286 & 0.2753 & 0.1517 & 0.2772 & 0.3908 \\
& PD~\cite{ZhangFHWSLZ21} & 0.3997 & 0.3543 & 0.3737 & 0.1433 & 0.2182 & 0.2606 & 0.1433 & 0.2622 & 0.3627 \\
& MACR~\cite{WeiFCWYH21} & 0.4176 & 0.3798 & 0.3973 & 0.1559 & 0.2389 & 0.2834 & 0.1559 & 0.2875 & 0.3870 \\
& CJMF$^{\dagger}$~\cite{ZhuHZC20} & 0.4093 & 0.3856 & 0.4097 & 0.1500 & 0.2408 & 0.2900 & 0.1500 & 0.2984 & 0.4075 \\
\cline{2-11}
& BISER (ours) & \textbf{0.4503}$^{*}$ & \textbf{0.4109}$^{*}$ & \textbf{0.4378}$^{**}$ & \textbf{0.1725}$^{*}$ & \textbf{0.2663}$^{**}$ & \textbf{0.3192}$^{**}$ & \textbf{0.1725}$^{*}$ & \textbf{0.3185}$^{*}$ & \textbf{0.4367}$^{**}$ \\
\cline{2-11}
& Gain (\%) & 10.03 & 6.56 & 6.85 & 14.99 & 10.58 & 10.08 & 14.99 & 6.74 & 7.16 \\
\hline
\multirow{10}{*}{Yahoo! R3}
& MF~\cite{HuKV08} & 0.1797 & 0.2081 & 0.2411 & 0.1071 & 0.1688 & 0.1970 & 0.1071 & 0.2225 & 0.3040 \\
& UAE~\cite{SedhainMSX15} & 0.1983 & 0.2235 & 0.2532 & 0.1198 & 0.1836 & 0.2104 & 0.1198 & 0.2362 & 0.3111 \\
& IAE~\cite{SedhainMSX15} & 0.2137 & 0.2355 & 0.2653 & 0.1309 & 0.1956 & 0.2232 & 0.1309 & 0.2461 & 0.3211 \\
\cline{2-11}
& RelMF~\cite{SaitoYNSN20} & 0.1837 & 0.2122 & 0.2453 & 0.1102 & 0.1728 & 0.2014 & 0.1102 & 0.2266 & 0.3080 \\
& AT~\cite{Saito20at} & 0.1912 & 0.2179 & 0.2506 & 0.1149 & 0.1786 & 0.2071 & 0.1149 & 0.2310 & 0.3125 \\
& PD~\cite{ZhangFHWSLZ21} & 0.1994 & 0.2308 & 0.2647 & 0.1211 & 0.1901 & 0.2207 & 0.1211 & 0.2459 & 0.3297 \\
& MACR~\cite{WeiFCWYH21} & 0.2044 & 0.2274 & 0.2571 & 0.1243 & 0.1882 & 0.2154 & 0.1243 & 0.2382 & 0.3133 \\
& CJMF$^{\dagger}$~\cite{ZhuHZC20} & 0.2151 & 0.2426 & 0.2715 & 0.1320 & 0.2018 & 0.2291 & 0.1320 & 0.2564 & 0.3297 \\
\cline{2-11}
& BISER (ours) & \textbf{0.2323}$^{**}$ & \textbf{0.2608}$^{**}$ & \textbf{0.2894}$^{**}$ & \textbf{0.1446}$^{**}$ & \textbf{0.2195}$^{**}$ & \textbf{0.2479}$^{**}$ & \textbf{0.1446}$^{**}$ & \textbf{0.2748}$^{**}$ & \textbf{0.3477}$^{**}$ \\
\cline{2-11}
& Gain (\%) & 7.99 & 7.52 & 6.60 & 9.58 & 8.77 & 8.20 & 9.58 & 7.18 & 5.47 \\
 
\bottomrule
\end{tabular}
\end{center}
\end{table*}

\begin{table}[t]
\centering
\caption{Accuracy comparison for the effect of self-inverse propensity weighting on Coat and Yahoo! R3 (\textbf{bold} font indicates the best model for accuracy). `+ Rel-IPW' indicates adopting the IPW method of \citet{SaitoYNSN20}, `+ Pre-SIPW' indicates adopting a propensity as a prediction of pre-trained model with biased click, and `+ SIPW' indicates adopting the self-inverse propensity weighting. The results are averaged over 10 and 5 runs on Coat and Yahoo! R3, respectively. Although we only report NDCG$@N$ results, we observe similar trends with the other metrics, MAP$@N$ and Recall$@N$.
}
\vspace{-3mm}
\label{tab:su}
\begin{center}
\renewcommand{\arraystretch}{0.9} 
\begin{tabular}{P{1.2cm}P{1.9cm}P{1.1cm}P{1.1cm}P{1.1cm}}
\toprule
 Datasets &   Models & NDCG$@$1 & NDCG$@$3 & NDCG$@$5 \\
\hline
\multirow{12}{*}{Coat}
 &  MF        & 0.3748          & 0.3441          & 0.3714          \\
 &  + Rel-IPW    & 0.3959          & 0.3659          & 0.3922          \\
 & + Pre-SIPW & \textbf{0.4162} & \textbf{0.3867} & \textbf{0.4120} \\
 & + SIPW  & 0.3993 & 0.3686 & 0.3963 \\
 \cline{2-5}
 &  UAE       & 0.3610          & 0.3546          & 0.3815          \\
 &  + Rel-IPW  & 0.3876          & 0.3720          & 0.4031          \\
 & + Pre-SIPW & 0.3990 & 0.3766 & 0.3982 \\
 &   + SIPW & \textbf{0.4262} & \textbf{0.3973} & \textbf{0.4259} \\
 \cline{2-5}
 &   IAE       & 0.3655          & 0.3560          & 0.3812          \\
 &   + Rel-IPW   & 0.3990          & 0.3716          & 0.3951          \\
 & + Pre-SIPW & 0.4286 & 0.3939 & \textbf{0.4183} \\
 &   + SIPW & \textbf{0.4334} & \textbf{0.3949} & 0.4172 \\
\hline
\multirow{12}{*}{Yahoo! R3}
 &  MF        & 0.1797          & 0.2081          & 0.2411          \\
 &  + Rel-IPW    & 0.1837          & 0.2122          & 0.2453          \\
 & + Pre-SIPW & 0.1870 & 0.2104 & 0.2405 \\
 &   + SIPW  & \textbf{0.1881} & \textbf{0.2176} & \textbf{0.2496} \\
 \cline{2-5}
 &   UAE       & 0.1983          & 0.2235          & 0.2532          \\
 &   + Rel-IPW   & 0.2054          & 0.2384          & 0.2715          \\
 & + Pre-SIPW & 0.2013 & 0.2256 & 0.2529 \\
 &  + SIPW & \textbf{0.2115} & \textbf{0.2461} & \textbf{0.2785} \\
 \cline{2-5}
 &   IAE       & 0.2137          & 0.2355          & 0.2653          \\
 &  + Rel-IPW   & 0.2146          & 0.2398          & 0.2708          \\
 & + Pre-SIPW & 0.2120 & 0.2360 & 0.2645 \\
 &  + SIPW & \textbf{0.2190} & \textbf{0.2504} & \textbf{0.2802} \\
\bottomrule
\end{tabular}
\end{center}
\vspace{-3mm}
\end{table}


\vspace{1mm}
\noindent
\textbf{Competing models}. We compare BISER with three conventional recommender models, \ie, MF~\cite{HuKV08}, UAE~\cite{SedhainMSX15}, and IAE~\cite{SedhainMSX15}, and five unbiased recommender models, \ie, RelMF~\cite{SaitoYNSN20}, AT~\cite{Saito20at}, CJMF~\cite{ZhuHZC20}, PD~\cite{ZhangFHWSLZ21}, and MACR~\cite{WeiFCWYH21}. Note that several methods \cite{SaitoYNSN20, Saito20at, ZhuHZC20, ZhangFHWSLZ21,WeiFCWYH21,ChenDQ0XCLY21,BonnerV18,ZhengGLHLJ21} have been proposed to remove the bias in implicit feedback, but some methods (\eg, \cite{ChenDQ0XCLY21,BonnerV18,ZhengGLHLJ21}) require MAR training data; therefore, they are not included.

\begin{itemize}[leftmargin=5mm]
\item \textbf{Matrix Factorization (MF)}~\cite{HuKV08}: The most popular recommender model for using linear factor embeddings in which the user preference score is predicted by a product of user- and item-embedding matrices.

\item \textbf{User-based AutoEncoder (UAE)}~\cite{SedhainMSX15}: This model learns non-linear item-item correlations for the users with a set of items using autoencoders.

\item \textbf{Item-based AutoEncoder (IAE)}~\cite{SedhainMSX15}: By transposing the rating matrix, this model learns non-linear user-user correlations, using autoencoders.

\item \textbf{Relevance Matrix Factorization (RelMF)}~\cite{SaitoYNSN20}: This utilizes the MNAR scenario for model training, effectively removing the bias caused by item popularity. The IPW-based model estimates the propensity scores using a heuristic function for item popularity.

\item \textbf{Asymmetric Tri-training (AT)}~\cite{Saito20at}: It uses two pre-trained models to generate reliable pseudo-ratings and the other model is trained with pseudo-ratings to make final predictions. Note that RelMF~\cite{SaitoYNSN20} is used for the pre-trained models.

\item \textbf{Combinational Joint Learning for Matrix Factorization (CJMF)}~\cite{ZhuHZC20}: As an IPW-based model, it introduces a joint training framework to estimate both unbiased relevance and unbiased propensity using multiple sub-models. For our datasets, we set the number of sub-models to eight. It is one of the most compelling competitors due to the performance advantage of unbiased recommender learning.

\item \textbf{Popularity-bias Deconfounding (PD)}~\cite{ZhangFHWSLZ21}: This uses a causal graph to remove confounding popularity bias through causal intervention. It is implemented using BPRMF~\cite{RendleFGS09}, as mentioned in~\cite{ZhangFHWSLZ21}.

\item \textbf{Model-Agnostic Counterfactual Reasoning (MACR)}~\cite{WeiFCWYH21}: This utilizes a cause-effect view by considering the direct effect of item properties on rank scores to remove the popularity bias. We report MACR using LightGCN~\cite{0001DWLZ020} because LightGCN is better than BPRMF~\cite{RendleFGS09}.
\end{itemize}

\vspace{1mm}
\noindent
\textbf{Evaluation metrics}. For \topN\ recommendation evaluation, we use three popular metrics: \emph{Normalized Discounted Cumulative Gain (NDCG@$N$)}, \emph{Mean Average Precision (MAP@$N$)}, and \emph{Recall@$N$}. We focus on measuring all metrics for the highest ranked items because predicting high-ranked items is critical. Considering the number of available ratings in the test set, we use $N = \{1, 3, 5\}$ for Coat and Yahoo! R3 datasets, while $N = \{10, 30, 50\}$ for ML-100K, ML-1M, ML-10M, and CiteULike datasets.

For MAR evaluation (Coat and Yahoo! R3), we adopt the \emph{Average-Over-All (AOA) evaluation}, which is the conventional metric that evenly normalizes the scores for all items. For the MNAR evaluation (MovieLens and CiteULike), we use both \emph{AOA} and \emph{unbiased evaluations}. \citet{YangCXWBE18} recently suggested an unbiased evaluation scheme and applies bias normalization to test scores. We follow the unbiased evaluation~\cite{YangCXWBE18} and use $\gamma = 2$ for the normalization parameter.

We employ AOA and unbiased evaluation metrics in~\cite{YangCXWBE18}:
\begin{equation}\label{eq:aoa_evaluation}
\hat{R}_{AOA}(\hat{Z}) = \frac{1}{|\mathcal{U}|}\sum_{u \in \mathcal{U}}{ \frac{1}{|\mathcal{S}_{u}|}\sum_{i \in \mathcal{S}_{u}}c(\hat{Z}_{ui})},
\end{equation}

\begin{equation}\label{eq:unbiased_evaluation}
\hat{R}_{unbiased}(\hat{Z}) = \frac{1}{|\mathcal{U}|} \sum_{u \in \mathcal{U}}{ \frac{1}{|\mathcal{S}_{u}|}   \sum_{i \in \mathcal{S}_{u}} \frac{c(\hat{Z}_{ui})}{P_{i}}},
\end{equation}
where $\mathcal{S}_{u}$ denotes a set of clicked item to user $u$ in the test set and $\hat{Z}_{ui}$ denotes the predicted ranking of item $i$ to user $u$. Also, the function $c(\cdot)$ denotes any \topN\ scoring metrics (\ie, NDCG$@N$, MAP$@N$, and Recall$@N$), and the propensity $P_{i}$ is calculated by the popularity of the item $i$. Because MNAR-MNAR evaluation does not have the ground-truth for unbiased test sets, it is difficult to measure the unbiased evaluation in eliminating the exposure bias of items. As an alternative, we evaluate whether the popularity bias is effectively eliminated, as done in~\cite{YangCXWBE18}, which is a more favorable setting for existing studies~\cite{SaitoYNSN20, Saito20at} that handle item popularity bias.

\vspace{1mm}
\noindent
\textbf{Reproducibility}. For all models, trainable weights are initialized using Xavier's method~\cite{GlorotB10}. Except for PD~\cite{ZhangFHWSLZ21} and MACR~\cite{WeiFCWYH21}, we optimized using Adagrad~\cite{DuchiHS10}, and for PD~\cite{ZhangFHWSLZ21} and MACR~\cite{WeiFCWYH21}, we optimized using Adam~\cite{KingmaB14}. For the MF-based models, the batch size is set to 1,024 ($2^{10}$) for Coat, Yahoo! R3, and MovieLens, and 16,384 ($2^{14}$) for CiteULike. Meanwhile, for the LightGCN-based model, the batch size is set to 2,048 and 8,192 for Coat and the remaining datasets, respectively. For the AE-based models, the batch size is set to one user/item by default. We performed a grid search to tune the latent dimension over \{50, 100, 200, 400\} for the AE-based models and \{32, 64, 128, 256\} for the MF- and LightGCN-based models. We also searched for the learning rate within [2e-1, 1e-5] and L2-regularization within the range of [1e-4, 1e-14] for each model. For the proposed BISER, we tuned two coefficients, $\lambda_{U}$ and $\lambda_{I}$, between \{0.1, 0.5, 0.9\}. Additionally, we set the maximum training epoch to 500. We also perform early stopping with five patience epochs for NDCG$@3$ and NDCG$@30$ in the MNAR-MAR and MNAR-MNAR settings, respectively. To implement RelMF~\cite{SaitoYNSN20}, CJMF~\cite{ZhuHZC20}, PD~\cite{ZhangFHWSLZ21}, and MACR~\cite{WeiFCWYH21}, we used the codes provided by each author. Our code and detailed hyperparameter settings are available at \url{https://github.com/Jaewoong-Lee/sigir_2022_BISER}.
\section{Experimental results}
\label{sec:result}

In this section, we report extensive experimental results for BISER in both the MNAR-MAR and MNAR-MNAR settings. We also analyze the performance of BISER by item group and the effectiveness of each component (\ie, SIPW and BU) in the MNAR-MAR setting. For the MNAR-MAR setting, we reasonably validate the effectiveness of BISER. For the MNAR-MNAR setting, we also evaluate whether BISER is still effective in general evaluation environments.

\subsection{MNAR-MAR Evaluation}

\noindent
\textbf{Performance comparison}. Table~\ref{tab:unbiased} reports the comparison results between BISER and the other competing models. From these results, we obtained several intriguing findings. First, BISER demonstrates significant and consistent performance gains across all metrics by 6.56-14.99\% on Coat and 5.47-9.58\% on Yahoo! R3, achieving state-of-the-art performance. This indicates that BISER is more effective in eliminating the bias than existing unbiased models because the high performance on MAR test sets indicates that the bias is successfully removed, or at least, our proposed model is more robust to bias than the competing models. Second, the unbiased models (RelMF, AT, PD, MACR, CJMF, and BISER) generally outperform the traditional models (UAE, IAE, and MF). This potentially implies that eliminating bias in the click data can improve the user experience. Finally, among the existing unbiased models, CJMF achieves better accuracy than other MF-based models due to an ensemble effect from eight sub-models and MACR~\cite{WeiFCWYH21} also shows competitive performance among the existing models.

\begin{figure}[t!]
\centering
\qquad\includegraphics[width=0.4\textwidth]{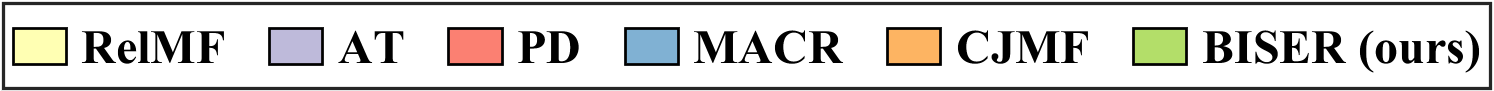}
\begin{tabular}{ccc}
\includegraphics[width=0.14\textwidth]{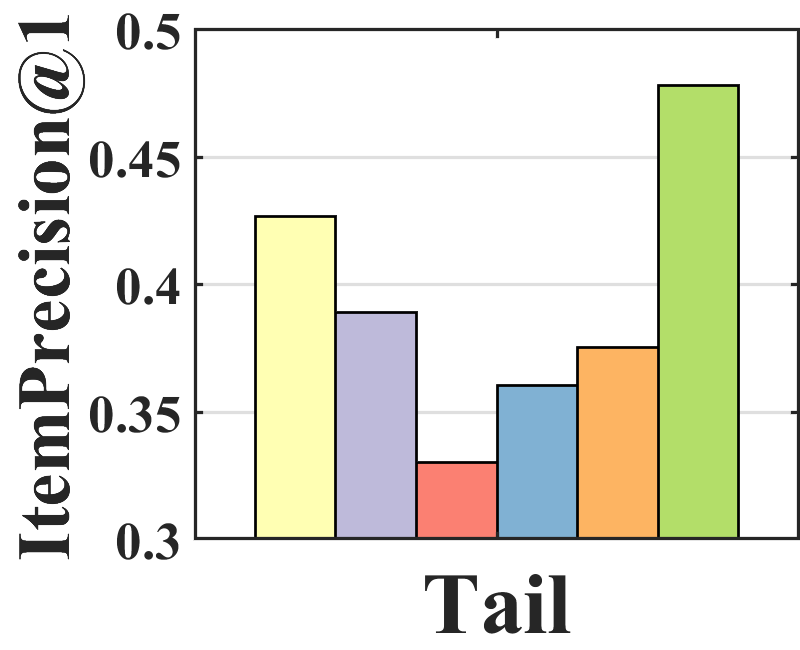} &
\includegraphics[width=0.14\textwidth]{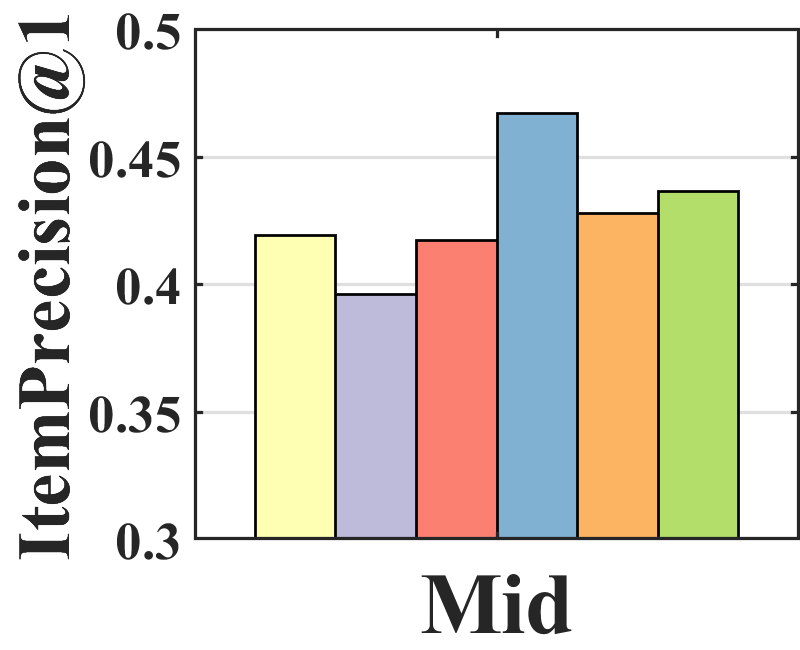} &
\includegraphics[width=0.14\textwidth]{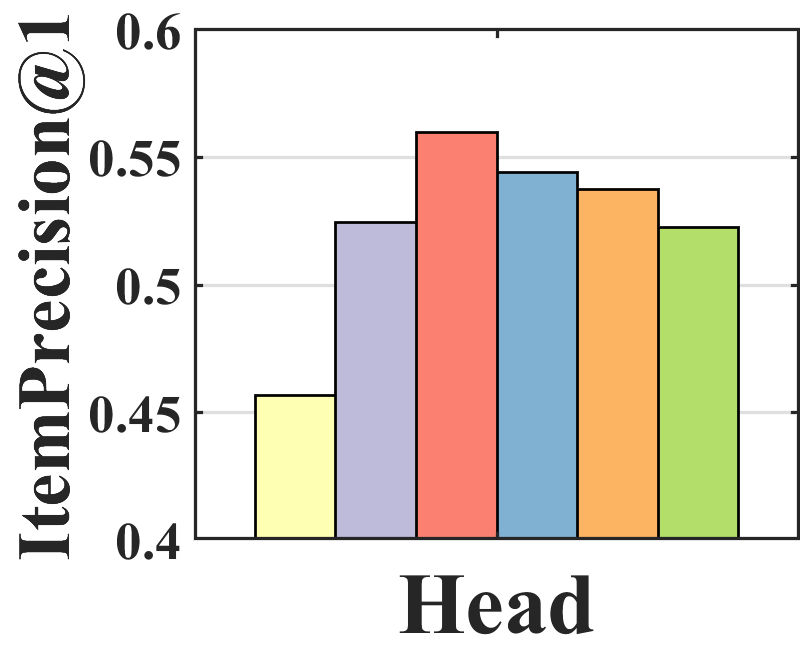} \\
\multicolumn{3}{c}{(a) Coat} 
\vspace{1mm} \\
\includegraphics[width=0.14\textwidth]{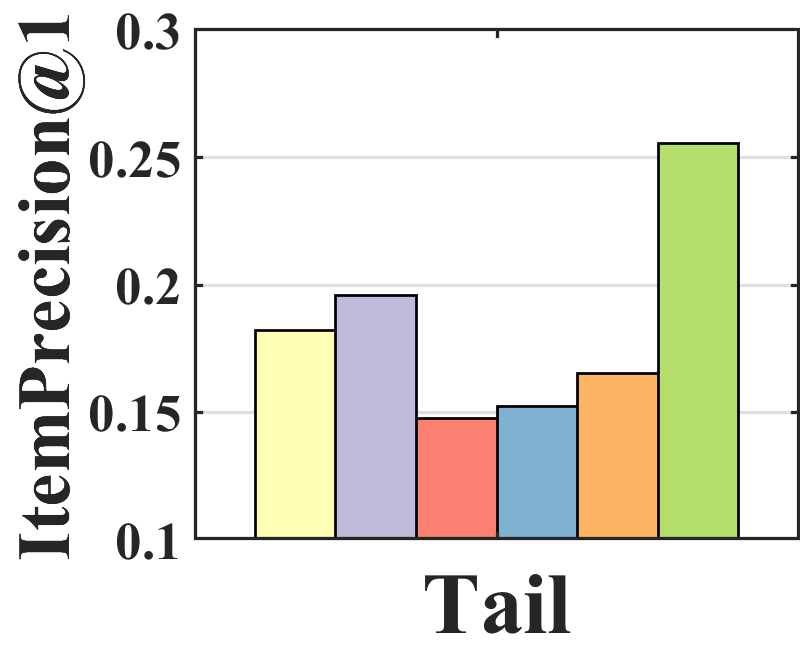} &
\includegraphics[width=0.14\textwidth]{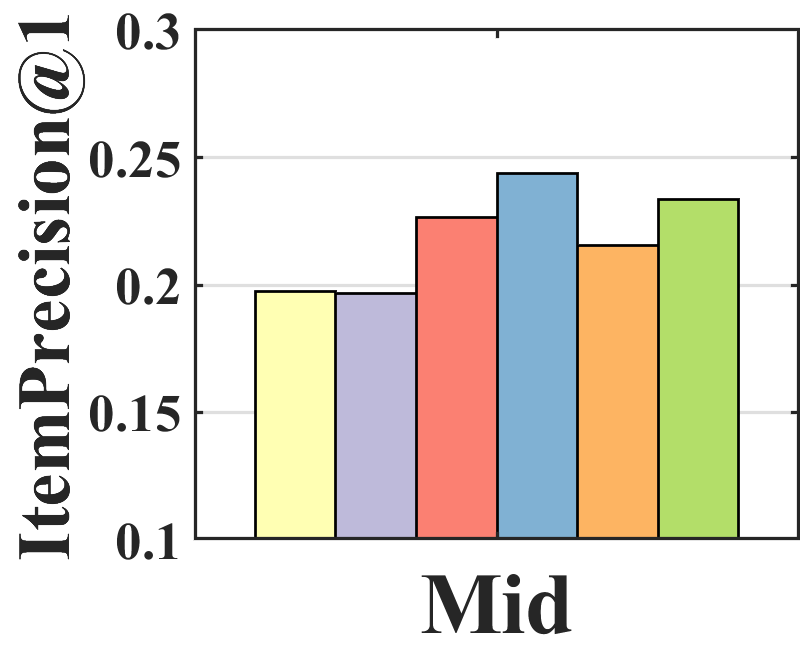} &
\includegraphics[width=0.14\textwidth]{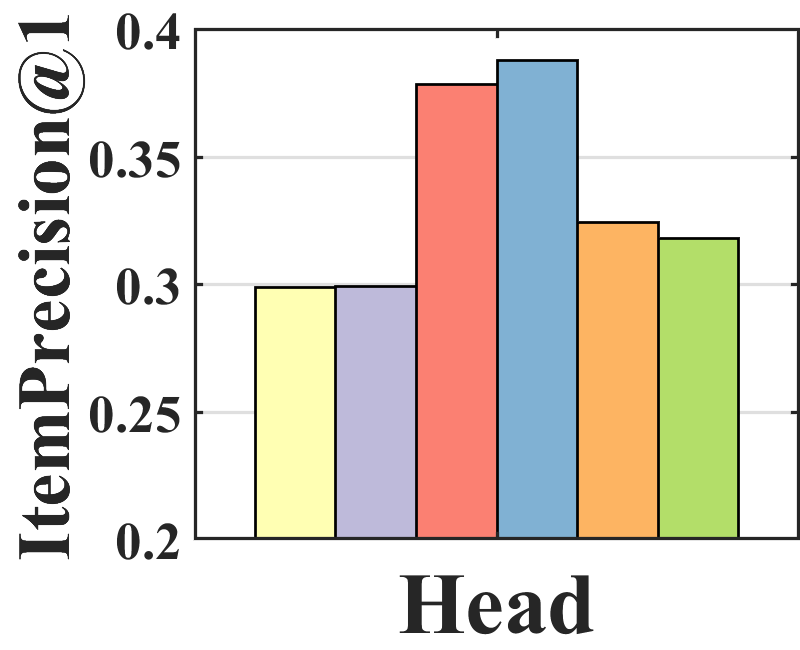} \\
\multicolumn{3}{c}{(b) Yahoo! R3}
\end{tabular}
\vskip -2mm
\caption{\emph{ItemPrecision$@$1} in different item groups on Coat and Yahoo! R3. All items are sorted by item popularity and divided into three groups with equal number of interactions. For Coat, tail, mid and head are the sets of items with 1-7, 7-12, and 12+ interactions, respectively. For each group, the number of items is 203, 66, and 31, respectively. For Yahoo! R3, tail, mid and head are the sets of items with 3-140, 140-700, and 700+ interactions, respectively, and the number of items for each group is 800, 168, and 32, respectively.}\label{fig:pop-performance}
\vskip -4mm
\end{figure}


\vspace{1mm}
\noindent
\textbf{Performance analysis}. To analyze where the performance improvement of BISER comes from, we investigate the performance of each item group according to item popularity. Specifically, we sort all items based on their popularity and split them into three groups (\ie, tail, mid, and head) with an equal number of interactions. Figure~\ref{fig:pop-performance} shows the average of \emph{ItemPrecision$@$1} for each item group. \emph{ItemPrecision$@$N} refers to the ratio of the items that appear in the relevance set to the \topN\ items for all users.
\begin{equation}\label{eq:item_precision}
ItemPrecision(i)@N =  \frac{ \sum_{u \in \mathcal{U}}{\mathbbm{1} \{ \hat{Z}_{ui} \leq  N \cap r_{ui} = 1\} }}{\sum_{u \in \mathcal{U}}{ \mathbbm{1} \{ \hat{Z}_{ui} \leq N \} }} .
\end{equation}




As depicted in Figure~\ref{fig:pop-performance}, each model shows a different performance trend in the item groups. BISER shows a significantly higher performance than the baselines in the tail group. This means that BISER correctly predicts at user's preference for unpopular items. RelMF~\cite{SaitoYNSN20} and AT~\cite{Saito20at} show similar performance over the three groups, whereas PD~\cite{ZhangFHWSLZ21} and MACR~\cite{WeiFCWYH21} show relatively high performance in the mid and head groups. This indicates that the performance gains of PD~\cite{ZhangFHWSLZ21} and MACR~\cite{WeiFCWYH21} mainly come from the popular item group. That is, BISER removes bias more effectively than other competing models, including the causal graph-based method. In addition, the higher the popularity (\ie, tail $\rightarrow$ mid $\rightarrow$ head), the higher is the overall performance of all the baselines. This shows that all the models more easily predict users' preferences for popular items than for unpopular items because popular items tend to be biased in training recommender models.

\vspace{1mm}
\noindent
\textbf{Effect of self-inverse propensity weighting (SIPW)}. To validate the effectiveness of SIPW, we compare four different groups of models: (i) na\"ive models (\ie, MF, UAE, and IAE); (ii) models adopting the IPW module from~\citet{SaitoYNSN20} (\ie, MF, UAE, and IAE + Rel-IPW); (iii) models with the pre-defined propensity as a prediction of the pre-trained model with biased click data (\ie, MF, UAE, and IAE + Pre-SIPW; a simple variant of SIPW); and (iv) models with SIPW (\ie, MF, UAE, and IAE + SIPW). Because of its model-agnostic properties, our poposed SIPW method can be applied to both MF- and AE-based models.

Table~\ref{tab:su} shows that the SIPW outperforms Rel-IPW by up to 9.96\% in Coat and up to 4.44\% in Yahoo! R3 over NDCG$@$1, 3, and 5. This suggests that SIPW helps eliminate the bias regardless of the backbone models. Additionally, the proposed method shows similar performance compared to pre-SIPW. This means that our approach of gradually removing bias without pre-training is efficient and effective in removing bias. In particular, the AE-based models have higher performance improvements than MF-based models. Because the AE-based models represent non-linearity by their modeling power, we can further enjoy high-performance gains.

\begin{figure}[t]
\centering
\includegraphics[width=0.45\textwidth]{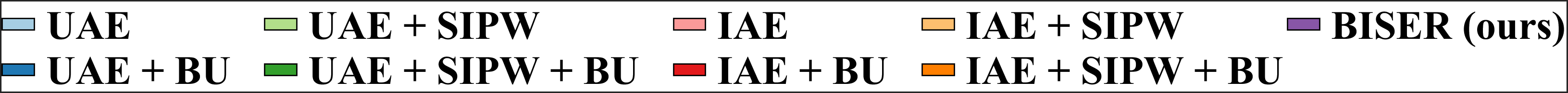}
\begin{tabular}{cc}
\renewcommand{\arraystretch}{0.8} 
\includegraphics[width=0.2\textwidth]{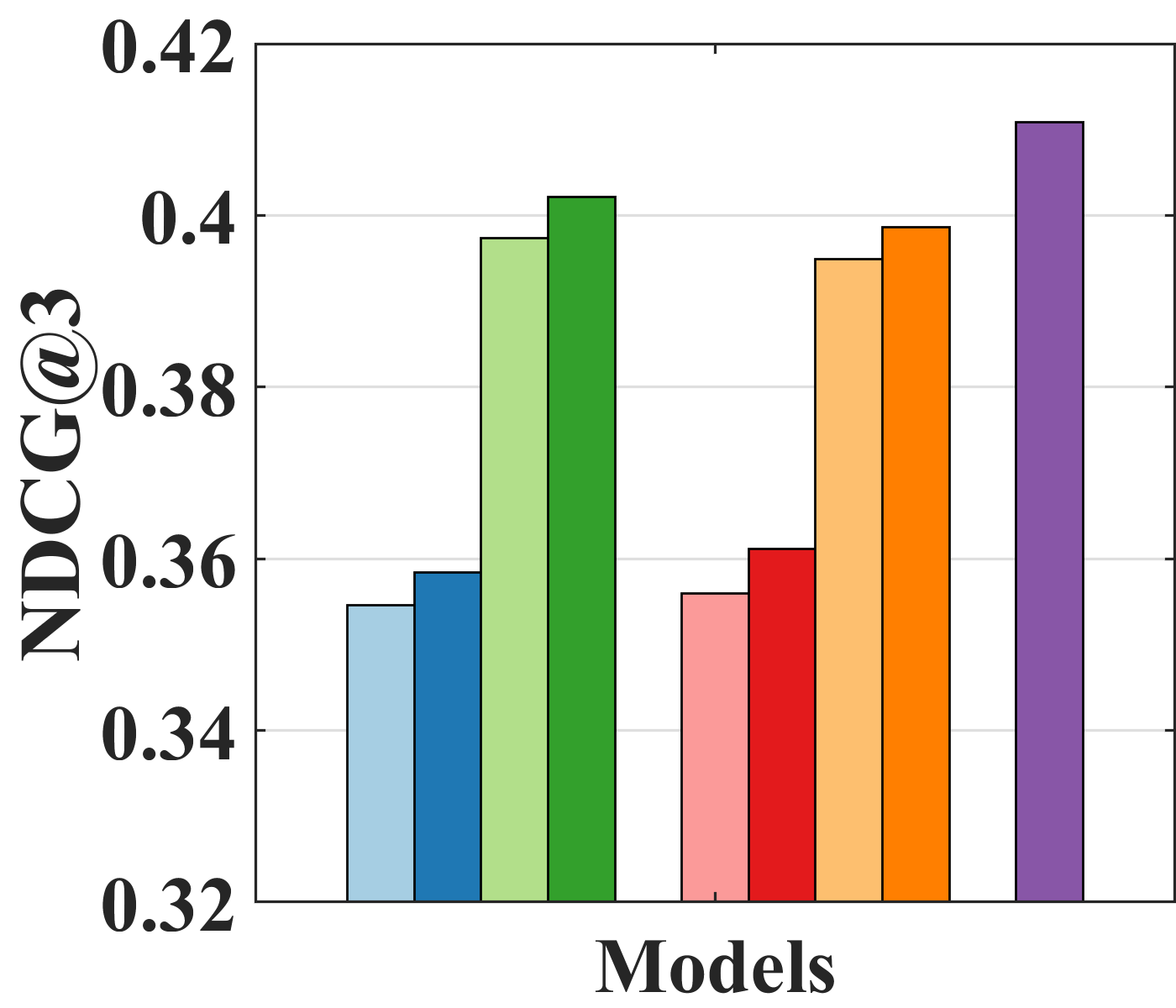} &
\includegraphics[width=0.2\textwidth]{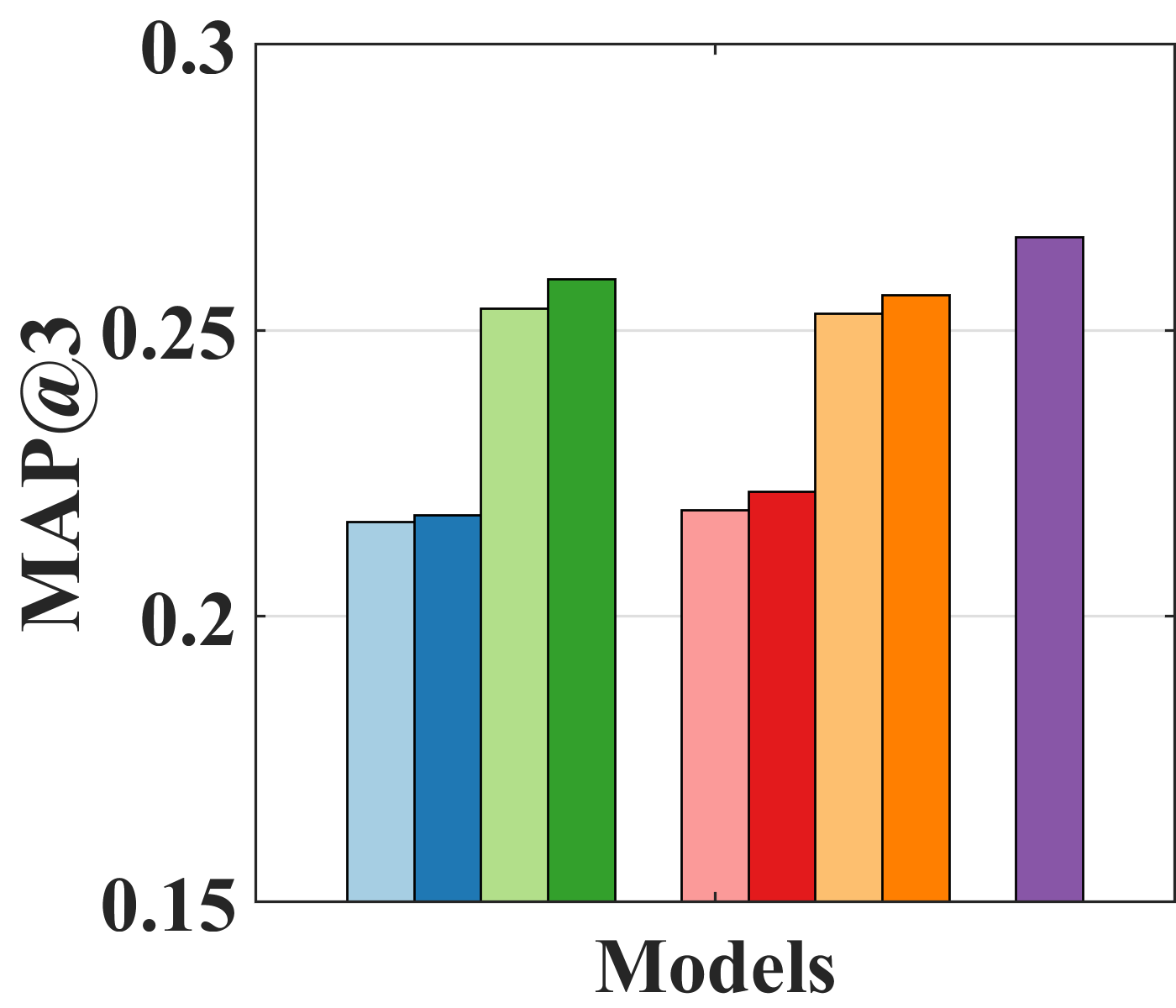} \vspace{-1mm}\\
\multicolumn{2}{c}{(a) Coat} \vspace{-1mm}\\ \\

\includegraphics[width=0.2\textwidth]{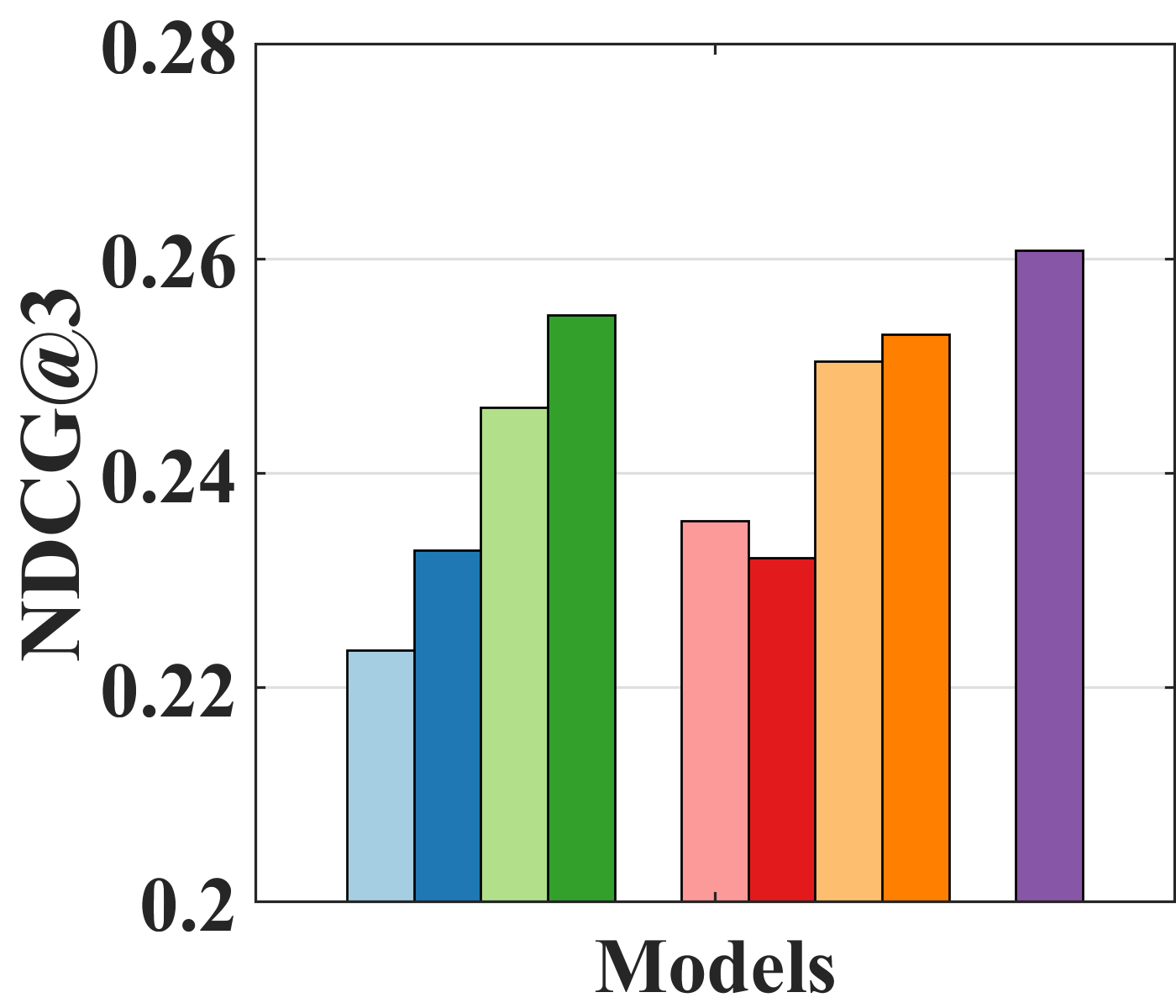} &
\includegraphics[width=0.2\textwidth]{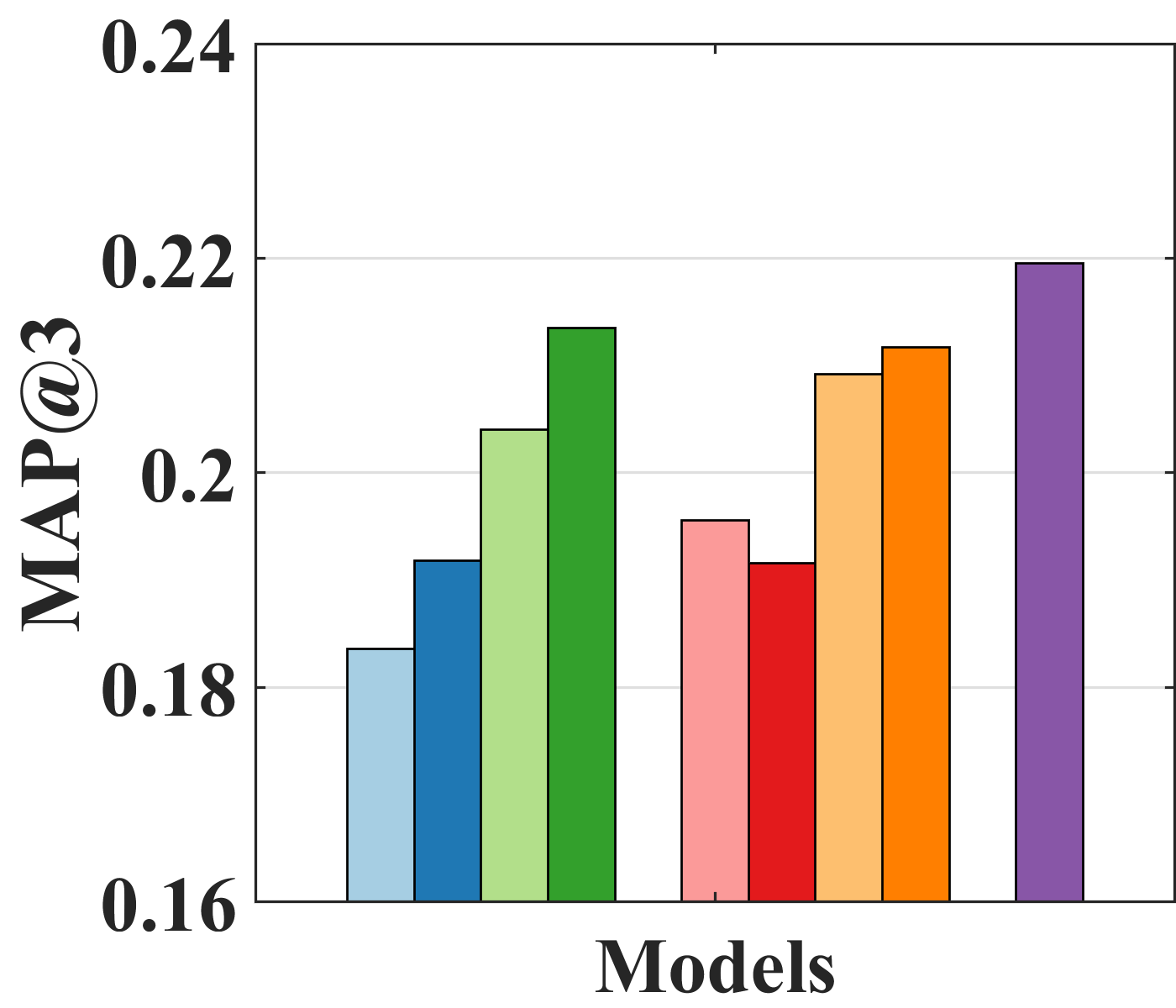} \vspace{-1mm}\\
\multicolumn{2}{c}{(b) Yahoo! R3}
\vspace{-2mm}
\end{tabular}
\caption{Effect of self-inverse propensity weighting (SIPW) and bilateral unbiased learning (BU) on Coat and Yahoo! R3. The suffix `+ SIPW' indicates adopting SIPW; the suffix `+ BU' indicates adopting BU; and the suffix `+ SIPW + BU' indicates adopting both SIPW and BU. The results are averaged over 10 and 5 runs on Coat and Yahoo! R3, respectively. We also observe similar trends with other metrics, $@1$ and $@5$.}\label{fig:cu}
\vspace{-2mm}
\end{figure}

\vspace{1mm}
\noindent
\textbf{Effect of bilateral unbiased learning (BU)}. We design four different baselines: (i) AE-based na\"ive models (UAE and IAE); (ii) models with BU (UAE and IAE + BU); (iii) models with SIPW (UAE and IAE + SIPW); and (iv) models with both SIPW and BU (UAE and IAE + SIPW + BU). Figure~\ref{fig:cu} indicates that adopting BU in all models, except for IAE on Yahoo! R3, improves performance. Specifically, IAE + SIPW + BU is improved by 4.11\% over IAE + SIPW on Coat and UAE + SIPW + BU over UAE + SIPW by 3.48\% on Yahoo! R3 in terms of NDCG$@$3. Although IAE slightly outperforms IAE + BU in Figure~\ref{fig:cu}(b), IAE + SIPW + BU outperforms IAE + SIPW on Yahoo! R3 by 2.61\% in terms of NDCG$@$3. Because BU is adopted in the na\"ive models without removing bias, the bias of IAE + BU could be intensified, which might worsen the performance.

\begin{figure}[t!]
\centering
\begin{tabular}{cc}
\renewcommand{\arraystretch}{0.8} 
\includegraphics[width=0.20\textwidth]{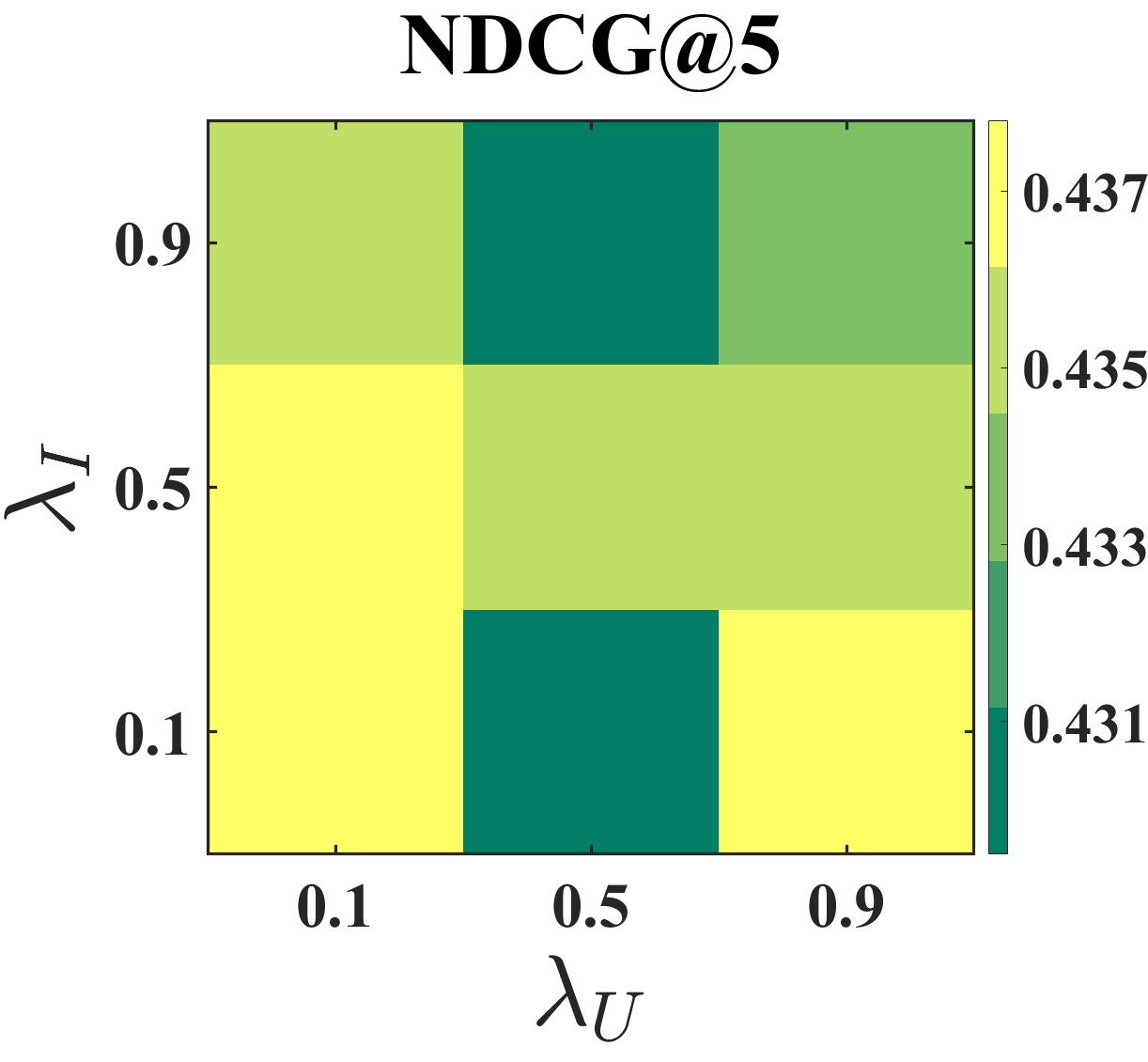} &
\includegraphics[width=0.20\textwidth]{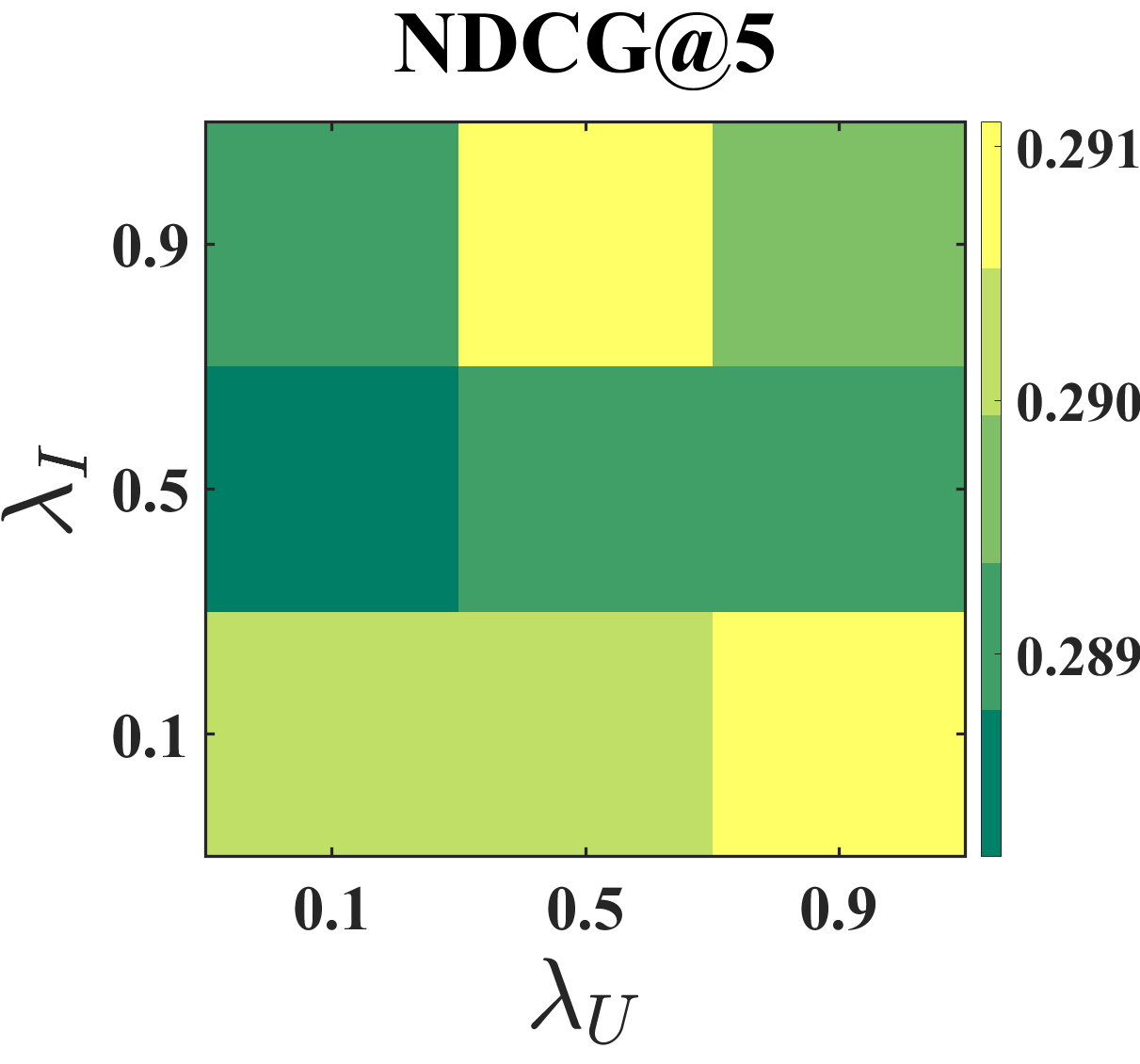} \\
\multicolumn{1}{c}{(a) Coat} &\multicolumn{1}{c}{(b) Yahoo! R3}
\vspace{-2mm}
\end{tabular}
\caption{Heatmap of hyperparameters $\lambda_U$ and $\lambda_I$ on Coat and Yahoo! R3. They are the coefficients in the proposed loss functions in Eqs.~\eqref{eq:proposed_final_1} and~\eqref{eq:proposed_final_2}. The results are averaged over 10 and 5 runs on Coat and Yahoo! R3, respectively.}\label{fig:heatmap}
\vspace{-4mm}
\end{figure}
\vspace{1mm}
\noindent
\textbf{Effect of coefficients $\lambda_U$ and $\lambda_I$}. 
Figure~\ref{fig:heatmap} shows the result of the grid search on Coat and Yahoo! R3 datasets with two coefficients $\lambda_U$ and $\lambda_I$ in the range \{0.1, 0.5, 0.9\}. Specifically, the NDCG$@5$ scores of BISER are between 0.4331--0.4378 in the Coat and 0.2890--0.2911 in the Yahoo! R3. For Coat, the proposed model tends to show high performances when $\lambda_U$ and $\lambda_I$ are relatively small values. For Yahoo! R3, our model tends to have high performance when $\lambda_U$ and $\lambda_I$ are large and small, respectively. Based on the performance trends, we set $\lambda_U = 0.1, \lambda_I = 0.5$ in Coat, and $\lambda_U = 0.9, \lambda_I = 0.1$ in Yahoo! R3. These results support the notion that the effects of SIPW and BU can vary depending on datasets.

\begin{figure*}[t]
\centering
\includegraphics[width=0.5\textwidth]{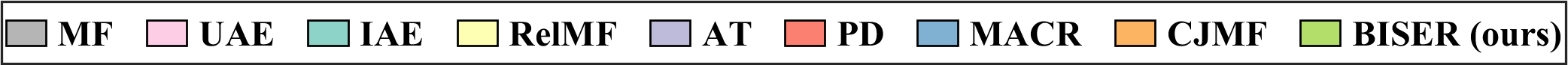}
\begin{tabular}{cccc}
\renewcommand{\arraystretch}{1.0} 
\includegraphics[width=0.22\textwidth]{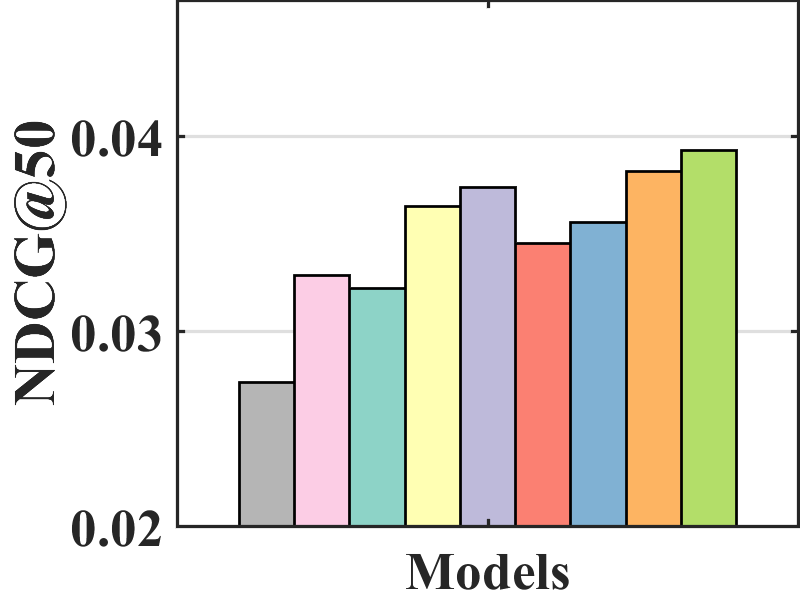} &
\includegraphics[width=0.22\textwidth]{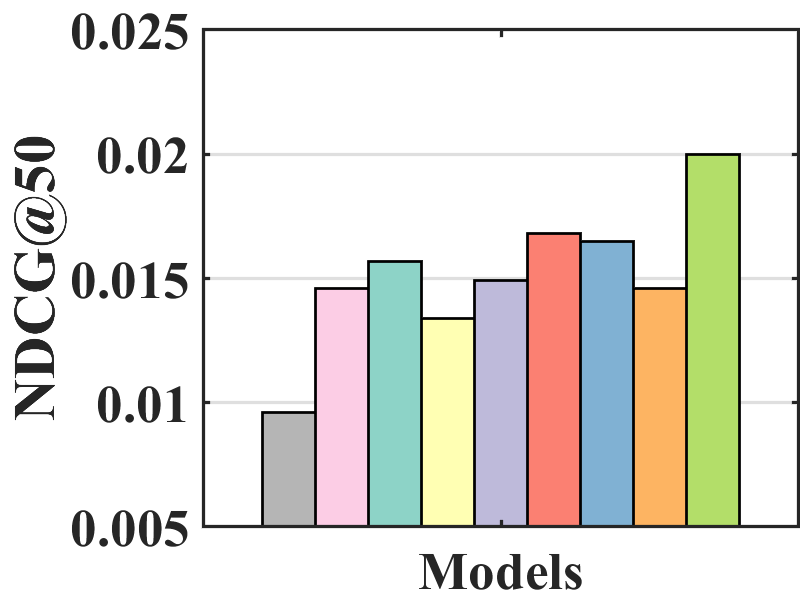} &
\includegraphics[width=0.22\textwidth]{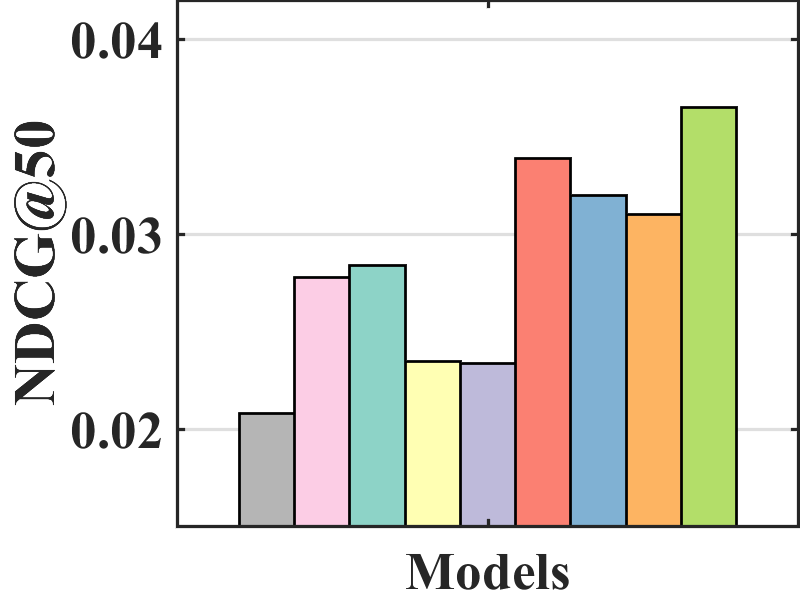} &
\includegraphics[width=0.22\textwidth]{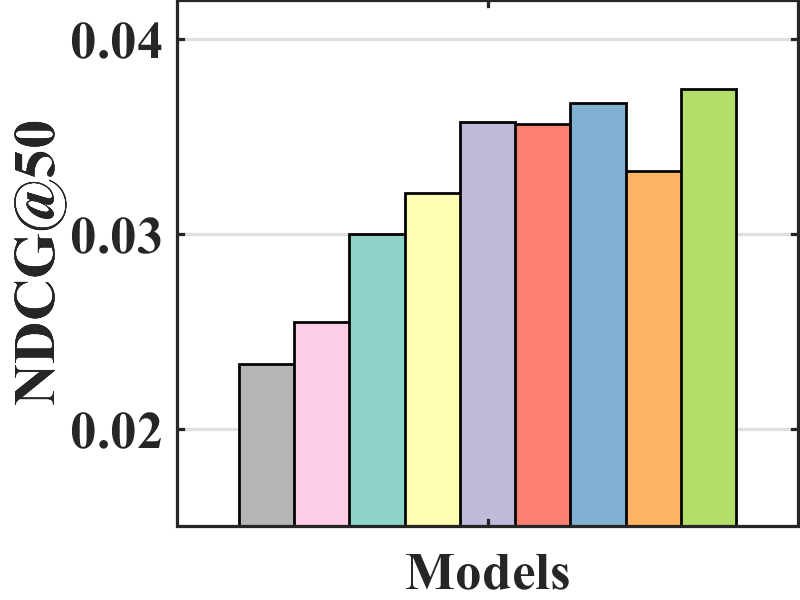} \\
\multicolumn{1}{c}{(a) ML-100K} & \multicolumn{1}{c}{(b) ML-1M} & \multicolumn{1}{c}{(c) ML-10M} & \multicolumn{1}{c}{(d) CiteULike}
\vspace{-2.5mm}
\end{tabular}
\caption{Accuracy comparison with unbiased evaluation. The results are averaged over 5 runs on ML-100K, ML-1M, ML-10M, and CiteULike. We also observe similar trends with other metrics, $@10$ and $@30$.}\label{fig:unbiased_eval} 


\vspace{3mm}

\centering
\includegraphics[width=0.5\textwidth]{Figures/legend_mnar.png}
\begin{tabular}{cccc}
\renewcommand{\arraystretch}{1.0} 
\includegraphics[width=0.22\textwidth]{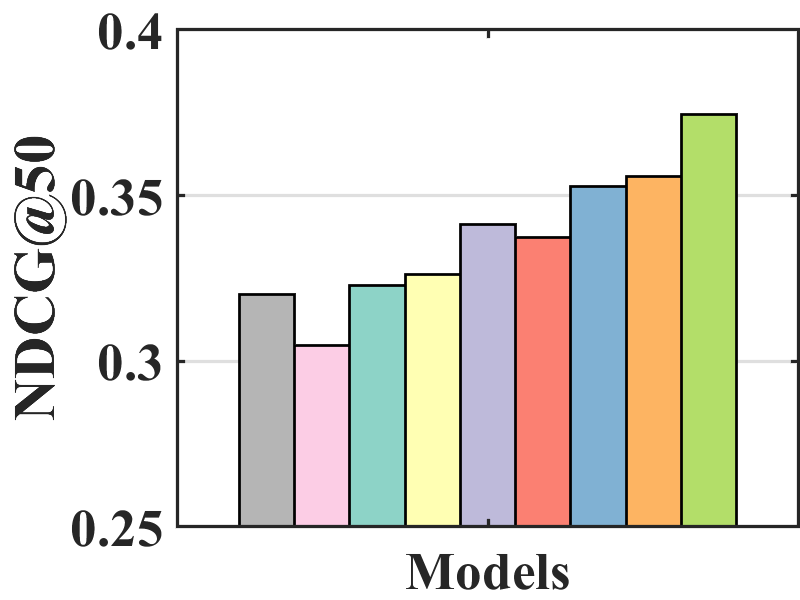} &
\includegraphics[width=0.22\textwidth]{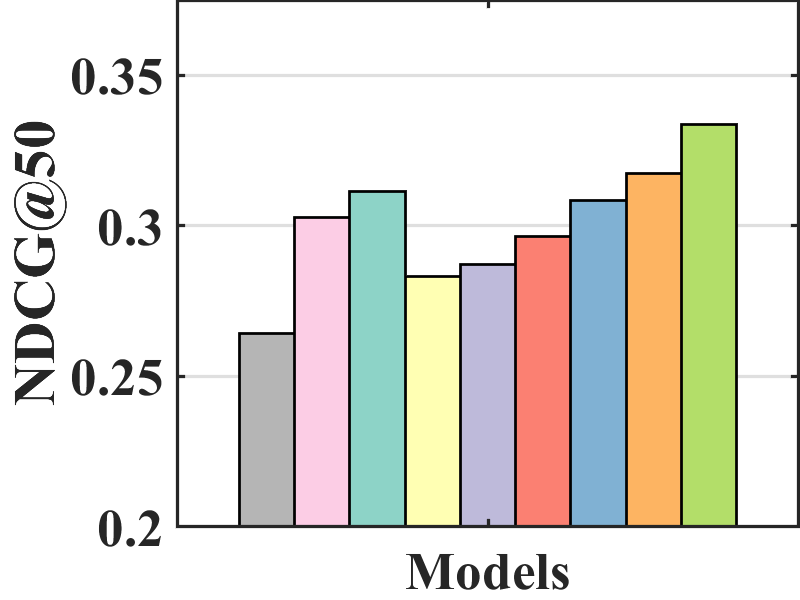} &
\includegraphics[width=0.22\textwidth]{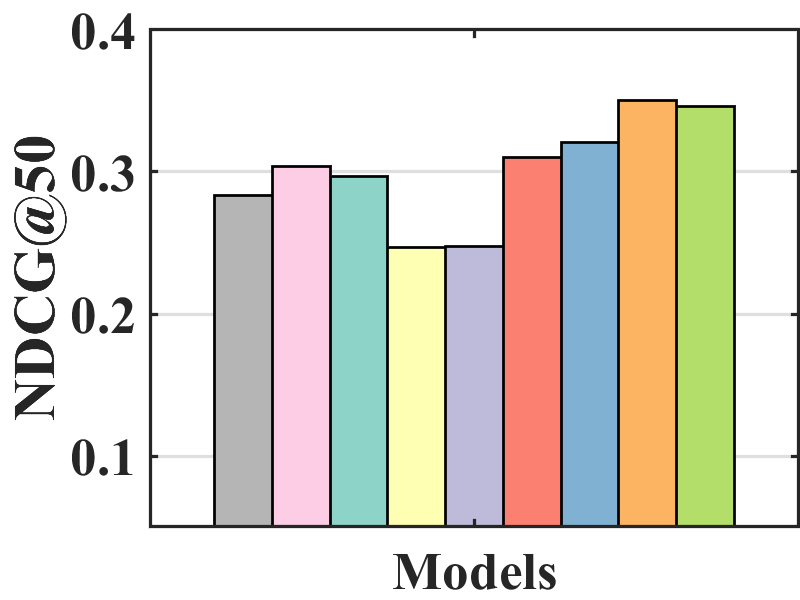} &
\includegraphics[width=0.22\textwidth]{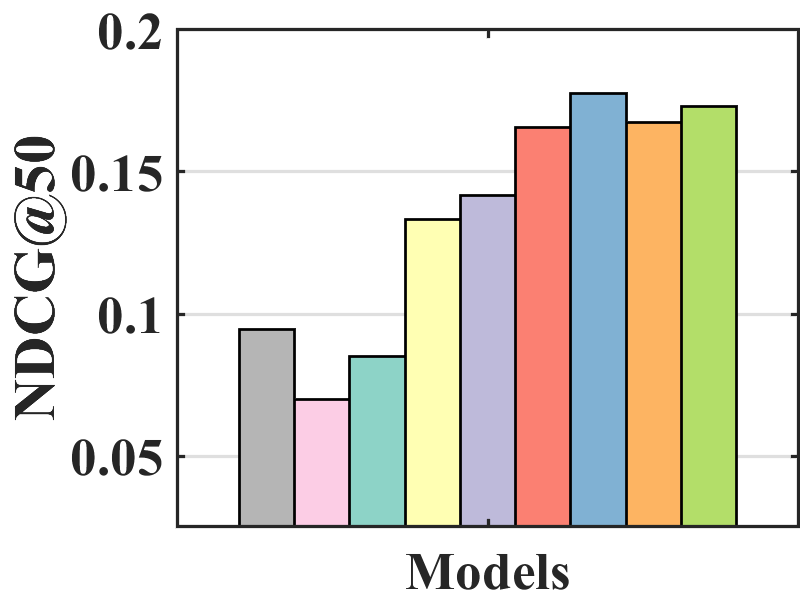} \\
\multicolumn{1}{c}{(a) ML-100K} & \multicolumn{1}{c}{(b) ML-1M} & \multicolumn{1}{c}{(c) ML-10M} & \multicolumn{1}{c}{(d) CiteULike}
\vspace{-2.5mm}
\end{tabular}
\caption{Accuracy comparison with AOA evaluation. The results are averaged over 5 runs on ML-100K, ML-1M, ML-10M, and CiteULike. We also observe similar trends with other metrics, $@10$ and $@30$.}\label{fig:aoa_eval} 
\vspace{-2mm}
\end{figure*}

\subsection{MNAR-MNAR Evaluation}
\noindent
\textbf{Performance comparison with unbiased evaluation}. In the MNAR-MNAR setting, the test set is biased. We use the unbiased evaluation proposed by~\citet{YangCXWBE18} to measure the debiasing effect on the biased test set.

Figure~\ref{fig:unbiased_eval} compares the performance of BISER with the baselines. We observe that BISER clearly and consistently outperforms the existing models. Specifically, BISER improved by 2.29, 19.32, 7.67, and 1.98\% over the second-best models (\ie, CJMF, PD, PD, and MACR) in ML-100K, ML-1M, ML-10M, and CiteULike, respectively, in terms of NDCG$@$50. This means that BISER effectively removes item bias regardless of data size and data type. Among the baselines, PD~\cite{ZhangFHWSLZ21} and MACR~\cite{WeiFCWYH21}, using causal graphs, generally exhibit higher performance than IPW-based methods (\ie, RelMF~\cite{SaitoYNSN20} and CJMF~\cite{ZhuHZC20}). This indicates that PD~\cite{ZhangFHWSLZ21} and MACR~\cite{WeiFCWYH21} eliminate popularity bias.

\vspace{1mm}
\noindent
\textbf{Performance comparison with AOA evaluation}. In the MNAR-MNAR setting, AOA evaluation is the conventional evaluation protocol. As depicted in Figure~\ref{fig:aoa_eval}, even in biased validation settings, BISER achieved a competitive performance over the other methods. Specifically, BISER performs best on the ML-100K and ML-1M datasets. In the ML-10M and CiteULike datasets, BISER shows the second-best model with a marginal difference from the best model. We conjecture that the success of BISER in MNAR-MNAR settings is possible by leveraging heterogeneous semantics using user- and item-based autoencoders via ensemble effects.

\section{Related Work}\label{sec:related}

Unbiased learning has been widely proposed for causal inference~\cite{WangBMN16, JoachimsSS17, WangGBMN18, AiBLGC18, HuWPL19, LeeSHLCL20, Wang0WW21, OvaisiAZVZ20, SchnabelSSCJ16, 0003ZS021, WangZSQ19, Saito20at, SaitoYNSN20, Saito20, ZhuHZC20, QinCMNQW20, BonnerV18, ZhangFHWSLZ21, WeiFCWYH21, ZhengGLHLJ21, JeonKLKL22} and missing data analyses~\cite{MarlinZRS07, Steck10, HernHG14, MarlinZ09, abs-1808-06581, MaC19, LiangCMB16}. It has since been adopted to bridge the gap between interaction and relevance data in the information retrieval (IR) community, that is, unbiased learning-to-rank (LTR). Unbiased recommender learning, inspired by unbiased LTR, has been actively studied to eliminate bias from explicit and implicit feedback.

\vspace{1mm}
\noindent
\textbf{Unbiased learning with explicit feedback.} Although explicit feedback provides both positive and negative samples, it is based on the MNAR assumption. To address this problem, existing studies are categorized into three types: \emph{imputation-based}, \emph{IPW-based}, and \emph{meta-learning-based methods}. First, imputation-based methods~\cite{MarlinZRS07, Steck10, HernHG14, MarlinZ09, abs-1808-06581, MaC19} estimate missing ratings to reduce the biased statistics of user ratings. \citet{Steck10} imputed a predefined constant value to all missing ratings. Some studies utilized various recommender models to impute different ratings: \citet{HernHG14} and \citet{abs-1808-06581} used an MF model; \citet{MarlinZRS07} used a multinomial mixture model, and \citet{MarlinZ09} used a conditional Bernoulli model. \citet{MaC19} adopted nuclear-norm-constrained matrix completion algorithm~\cite{abs-1209-3672} for rating imputation. Second, IPW-based methods~\cite{SchnabelSSCJ16, 0003ZS021, WangZSQ19} estimate propensity scores for user-item ratings. \citet{SchnabelSSCJ16} first adopted the IPW method and predicted the propensity score using Naive Bayes or logistic regression. \citet{0003ZS021} trained a propensity estimator with few unbiased ratings. In addition, \citet{WangZSQ19} combined IPW and imputation methods to eliminate bias and take advantage of both methods. Recently, a meta-learning-based method~\cite{Saito20at} was adopted to overcome the high-variance problem of IPW. \citet{Saito20at} trained a model using the output of one of multiple models as unbiased pseudo-labels. 


\vspace{1mm}
\noindent
\textbf{Unbiased learning with implicit feedback.} Unlike explicit feedback, implicit feedback provides only observed feedback. It is crucial to design an exposure bias for these items. Specifically, exposure bias can be considered for both user- and model-oriented reasons. Because users tend to recognize popular items, user clicks are biased towards popular items, \ie, popularity bias. To eliminate the popularity bias, \citet{LiangCMB16} determined the weight based on the prior probability of exposure. In addition, \cite{SaitoYNSN20, Saito20, ZhuHZC20, QinCMNQW20, LeePL21} used an IPW-based method to remove bias. While \cite{SaitoYNSN20, Saito20} simply utilized the number of ratings per item (\ie, item popularity) as the propensity score, \citet{ZhuHZC20} used an additional model to estimate the propensity score. \citet{QinCMNQW20} also introduced a propensity estimator that utilized additional attributes related to the recommender model (\eg, user interface of the application and type of recommender model). \citet{LeePL21} introduced a propensity score to both unclicked and clicked user feedback. \citet{ChenDQ0XCLY21} used small unbiased feedback to eliminate bias. Additionally, causal embedding- and graph-based training methods~\cite{BonnerV18, ZhengGLHLJ21, WeiFCWYH21, ZhangFHWSLZ21} have been introduced to overcome the sensitivity of IPW strategies. \citet{BonnerV18} introduced causal embedding using small uniformly collected feedback. \citet{ZhengGLHLJ21} introduced causal embedding and causal graph to disentangle user interests and conformity with small unbiased feedback. \citet{WeiFCWYH21} designed a causal graph with the item and user priors and \citet{ZhangFHWSLZ21} used causal intervention using a causal graph to eliminate bias. In contrast, our proposed method eliminates exposure bias caused by recommender models without additional assumptions for causality and model training.

 \section{Conclusion}\label{sec:conclusion}

This paper proposes a novel unbiased recommender model, namely \emph{BIlateral SElf-unbiased Recommender learning (BISER)}. To the best of our knowledge, this is the first paper to introduce self-inverse propensity weighting to eliminate exposure bias of items during model training. Then, we employed bilateral learning that takes advantage of user- and item-based autoencoders with heterogeneous information, capturing different hidden correlations across users/items. This process helped alleviate high variance in the estimated inverse propensity scores. Extensive experiments demonstrated that the BISER consistently outperformed existing unbiased recommender models in two evaluation protocols: MNAR-MAR and MNAR-MNAR settings.

\bibliographystyle{ACM-Reference-Format}
\bibliography{references}


\begin{thebibliography}{64}


\ifx \showCODEN    \undefined \def \showCODEN     #1{\unskip}     \fi
\ifx \showDOI      \undefined \def \showDOI       #1{#1}\fi
\ifx \showISBNx    \undefined \def \showISBNx     #1{\unskip}     \fi
\ifx \showISBNxiii \undefined \def \showISBNxiii  #1{\unskip}     \fi
\ifx \showISSN     \undefined \def \showISSN      #1{\unskip}     \fi
\ifx \showLCCN     \undefined \def \showLCCN      #1{\unskip}     \fi
\ifx \shownote     \undefined \def \shownote      #1{#1}          \fi
\ifx \showarticletitle \undefined \def \showarticletitle #1{#1}   \fi
\ifx \showURL      \undefined \def \showURL       {\relax}        \fi
\providecommand\bibfield[2]{#2}
\providecommand\bibinfo[2]{#2}
\providecommand\natexlab[1]{#1}
\providecommand\showeprint[2][]{arXiv:#2}

\bibitem[\protect\citeauthoryear{Adomavicius and Tuzhilin}{Adomavicius and
  Tuzhilin}{2005}]%
        {AdomaviciusT05}
\bibfield{author}{\bibinfo{person}{Gediminas Adomavicius} {and}
  \bibinfo{person}{Alexander Tuzhilin}.} \bibinfo{year}{2005}\natexlab{}.
\newblock \showarticletitle{Toward the Next Generation of Recommender Systems:
  {A} Survey of the State-of-the-Art and Possible Extensions}.
\newblock \bibinfo{journal}{\emph{{IEEE} Trans. Knowl. Data Eng.}}
  \bibinfo{volume}{17}, \bibinfo{number}{6} (\bibinfo{year}{2005}),
  \bibinfo{pages}{734--749}.
\newblock


\bibitem[\protect\citeauthoryear{Ai, Bi, Luo, Guo, and Croft}{Ai
  et~al\mbox{.}}{2018}]%
        {AiBLGC18}
\bibfield{author}{\bibinfo{person}{Qingyao Ai}, \bibinfo{person}{Keping Bi},
  \bibinfo{person}{Cheng Luo}, \bibinfo{person}{Jiafeng Guo}, {and}
  \bibinfo{person}{W.~Bruce Croft}.} \bibinfo{year}{2018}\natexlab{}.
\newblock \showarticletitle{Unbiased Learning to Rank with Unbiased Propensity
  Estimation}. In \bibinfo{booktitle}{\emph{SIGIR}}. \bibinfo{pages}{385--394}.
\newblock


\bibitem[\protect\citeauthoryear{Azizzadenesheli, Liu, Yang, and
  Anandkumar}{Azizzadenesheli et~al\mbox{.}}{2019}]%
        {Azizzadenesheli19}
\bibfield{author}{\bibinfo{person}{Kamyar Azizzadenesheli},
  \bibinfo{person}{Anqi Liu}, \bibinfo{person}{Fanny Yang}, {and}
  \bibinfo{person}{Animashree Anandkumar}.} \bibinfo{year}{2019}\natexlab{}.
\newblock \showarticletitle{Regularized Learning for Domain Adaptation under
  Label Shifts}. In \bibinfo{booktitle}{\emph{{ICLR} (Poster)}}.
  \bibinfo{publisher}{OpenReview.net}.
\newblock


\bibitem[\protect\citeauthoryear{Bonner and Vasile}{Bonner and Vasile}{2018}]%
        {BonnerV18}
\bibfield{author}{\bibinfo{person}{Stephen Bonner} {and}
  \bibinfo{person}{Flavian Vasile}.} \bibinfo{year}{2018}\natexlab{}.
\newblock \showarticletitle{Causal embeddings for recommendation}. In
  \bibinfo{booktitle}{\emph{RecSys}}. \bibinfo{publisher}{{ACM}},
  \bibinfo{pages}{104--112}.
\newblock


\bibitem[\protect\citeauthoryear{Chen, Dong, Qiu, He, Xin, Chen, Lin, and
  Yang}{Chen et~al\mbox{.}}{2021}]%
        {ChenDQ0XCLY21}
\bibfield{author}{\bibinfo{person}{Jiawei Chen}, \bibinfo{person}{Hande Dong},
  \bibinfo{person}{Yang Qiu}, \bibinfo{person}{Xiangnan He},
  \bibinfo{person}{Xin Xin}, \bibinfo{person}{Liang Chen},
  \bibinfo{person}{Guli Lin}, {and} \bibinfo{person}{Keping Yang}.}
  \bibinfo{year}{2021}\natexlab{}.
\newblock \showarticletitle{AutoDebias: Learning to Debias for Recommendation}.
  In \bibinfo{booktitle}{\emph{{SIGIR}}}. \bibinfo{pages}{21--30}.
\newblock


\bibitem[\protect\citeauthoryear{Chen, Dong, Wang, Feng, Wang, and He}{Chen
  et~al\mbox{.}}{2020}]%
        {ChenDWFWH20}
\bibfield{author}{\bibinfo{person}{Jiawei Chen}, \bibinfo{person}{Hande Dong},
  \bibinfo{person}{Xiang Wang}, \bibinfo{person}{Fuli Feng},
  \bibinfo{person}{Meng Wang}, {and} \bibinfo{person}{Xiangnan He}.}
  \bibinfo{year}{2020}\natexlab{}.
\newblock \showarticletitle{Bias and Debias in Recommender System: {A} Survey
  and Future Directions}.
\newblock \bibinfo{journal}{\emph{CoRR}}  \bibinfo{volume}{abs/2010.03240}
  (\bibinfo{year}{2020}).
\newblock


\bibitem[\protect\citeauthoryear{Choi, Jeong, Lee, and Lee}{Choi
  et~al\mbox{.}}{2021}]%
        {ChoiJLL21}
\bibfield{author}{\bibinfo{person}{Minjin Choi}, \bibinfo{person}{Yoonki
  Jeong}, \bibinfo{person}{Joonseok Lee}, {and} \bibinfo{person}{Jongwuk Lee}.}
  \bibinfo{year}{2021}\natexlab{}.
\newblock \showarticletitle{Local Collaborative Autoencoders}. In
  \bibinfo{booktitle}{\emph{WSDM}}. \bibinfo{pages}{734--742}.
\newblock


\bibitem[\protect\citeauthoryear{Choi, Kim, Lee, Shim, and Lee}{Choi
  et~al\mbox{.}}{2022}]%
        {ChoiKLSL22}
\bibfield{author}{\bibinfo{person}{Minjin Choi}, \bibinfo{person}{Jinhong Kim},
  \bibinfo{person}{Joonsek Lee}, \bibinfo{person}{Hyunjung Shim}, {and}
  \bibinfo{person}{Jongwuk Lee}.} \bibinfo{year}{2022}\natexlab{}.
\newblock \showarticletitle{S-Walk: Accurate and Scalable Session-based
  Recommendationwith Random Walks}. In \bibinfo{booktitle}{\emph{WSDM}}.
\newblock


\bibitem[\protect\citeauthoryear{Davenport, Plan, van~den Berg, and
  Wootters}{Davenport et~al\mbox{.}}{2012}]%
        {abs-1209-3672}
\bibfield{author}{\bibinfo{person}{Mark~A. Davenport}, \bibinfo{person}{Yaniv
  Plan}, \bibinfo{person}{Ewout van~den Berg}, {and} \bibinfo{person}{Mary
  Wootters}.} \bibinfo{year}{2012}\natexlab{}.
\newblock \showarticletitle{1-Bit Matrix Completion}.
\newblock \bibinfo{journal}{\emph{CoRR}}  \bibinfo{volume}{abs/1209.3672}
  (\bibinfo{year}{2012}).
\newblock


\bibitem[\protect\citeauthoryear{Duchi, Hazan, and Singer}{Duchi
  et~al\mbox{.}}{2010}]%
        {DuchiHS10}
\bibfield{author}{\bibinfo{person}{John~C. Duchi}, \bibinfo{person}{Elad
  Hazan}, {and} \bibinfo{person}{Yoram Singer}.}
  \bibinfo{year}{2010}\natexlab{}.
\newblock \showarticletitle{Adaptive Subgradient Methods for Online Learning
  and Stochastic Optimization}. In \bibinfo{booktitle}{\emph{COLT}}.
  \bibinfo{pages}{257--269}.
\newblock


\bibitem[\protect\citeauthoryear{Gilotte, Calauz{\`{e}}nes, Nedelec, Abraham,
  and Doll{\'{e}}}{Gilotte et~al\mbox{.}}{2018}]%
        {GilotteCNAD18}
\bibfield{author}{\bibinfo{person}{Alexandre Gilotte},
  \bibinfo{person}{Cl{\'{e}}ment Calauz{\`{e}}nes}, \bibinfo{person}{Thomas
  Nedelec}, \bibinfo{person}{Alexandre Abraham}, {and} \bibinfo{person}{Simon
  Doll{\'{e}}}.} \bibinfo{year}{2018}\natexlab{}.
\newblock \showarticletitle{Offline {A/B} Testing for Recommender Systems}. In
  \bibinfo{booktitle}{\emph{{WSDM}}}. \bibinfo{pages}{198--206}.
\newblock


\bibitem[\protect\citeauthoryear{Glorot and Bengio}{Glorot and Bengio}{2010}]%
        {GlorotB10}
\bibfield{author}{\bibinfo{person}{Xavier Glorot} {and} \bibinfo{person}{Yoshua
  Bengio}.} \bibinfo{year}{2010}\natexlab{}.
\newblock \showarticletitle{Understanding the difficulty of training deep
  feedforward neural networks}. In \bibinfo{booktitle}{\emph{AISTATS}},
  Vol.~\bibinfo{volume}{9}. \bibinfo{pages}{249--256}.
\newblock


\bibitem[\protect\citeauthoryear{He, Deng, Wang, Li, Zhang, and Wang}{He
  et~al\mbox{.}}{2020}]%
        {0001DWLZ020}
\bibfield{author}{\bibinfo{person}{Xiangnan He}, \bibinfo{person}{Kuan Deng},
  \bibinfo{person}{Xiang Wang}, \bibinfo{person}{Yan Li},
  \bibinfo{person}{Yong{-}Dong Zhang}, {and} \bibinfo{person}{Meng Wang}.}
  \bibinfo{year}{2020}\natexlab{}.
\newblock \showarticletitle{LightGCN: Simplifying and Powering Graph
  Convolution Network for Recommendation}. In
  \bibinfo{booktitle}{\emph{SIGIR}}. \bibinfo{pages}{639--648}.
\newblock


\bibitem[\protect\citeauthoryear{He, Du, Wang, Tian, Tang, and Chua}{He
  et~al\mbox{.}}{2018}]%
        {HeDWTTC18}
\bibfield{author}{\bibinfo{person}{Xiangnan He}, \bibinfo{person}{Xiaoyu Du},
  \bibinfo{person}{Xiang Wang}, \bibinfo{person}{Feng Tian},
  \bibinfo{person}{Jinhui Tang}, {and} \bibinfo{person}{Tat{-}Seng Chua}.}
  \bibinfo{year}{2018}\natexlab{}.
\newblock \showarticletitle{Outer Product-based Neural Collaborative
  Filtering}. In \bibinfo{booktitle}{\emph{IJCAI}}.
  \bibinfo{pages}{2227--2233}.
\newblock


\bibitem[\protect\citeauthoryear{He, Liao, Zhang, Nie, Hu, and Chua}{He
  et~al\mbox{.}}{2017}]%
        {HeLZNHC17}
\bibfield{author}{\bibinfo{person}{Xiangnan He}, \bibinfo{person}{Lizi Liao},
  \bibinfo{person}{Hanwang Zhang}, \bibinfo{person}{Liqiang Nie},
  \bibinfo{person}{Xia Hu}, {and} \bibinfo{person}{Tat{-}Seng Chua}.}
  \bibinfo{year}{2017}\natexlab{}.
\newblock \showarticletitle{Neural Collaborative Filtering}. In
  \bibinfo{booktitle}{\emph{WWW}}. \bibinfo{pages}{173--182}.
\newblock


\bibitem[\protect\citeauthoryear{Hernández-Lobato, Houlsby, and
  Ghahramani}{Hernández-Lobato et~al\mbox{.}}{2014}]%
        {HernHG14}
\bibfield{author}{\bibinfo{person}{J.M. Hernández-Lobato},
  \bibinfo{person}{Neil Houlsby}, {and} \bibinfo{person}{Z. Ghahramani}.}
  \bibinfo{year}{2014}\natexlab{}.
\newblock \showarticletitle{Probabilistic Matrix Factorization with Non-Random
  Missing Data}. In \bibinfo{booktitle}{\emph{ICML}}.
  \bibinfo{pages}{II–1512–II–1520}.
\newblock


\bibitem[\protect\citeauthoryear{Hu, Koren, and Volinsky}{Hu
  et~al\mbox{.}}{2008}]%
        {HuKV08}
\bibfield{author}{\bibinfo{person}{Yifan Hu}, \bibinfo{person}{Yehuda Koren},
  {and} \bibinfo{person}{Chris Volinsky}.} \bibinfo{year}{2008}\natexlab{}.
\newblock \showarticletitle{Collaborative Filtering for Implicit Feedback
  Datasets}. In \bibinfo{booktitle}{\emph{ICDM}}. \bibinfo{pages}{263--272}.
\newblock


\bibitem[\protect\citeauthoryear{Hu, Wang, Peng, and Li}{Hu
  et~al\mbox{.}}{2019}]%
        {HuWPL19}
\bibfield{author}{\bibinfo{person}{Ziniu Hu}, \bibinfo{person}{Yang Wang},
  \bibinfo{person}{Qu Peng}, {and} \bibinfo{person}{Hang Li}.}
  \bibinfo{year}{2019}\natexlab{}.
\newblock \showarticletitle{Unbiased LambdaMART: An Unbiased Pairwise
  Learning-to-Rank Algorithm}. In \bibinfo{booktitle}{\emph{WWW}}.
  \bibinfo{pages}{2830--2836}.
\newblock


\bibitem[\protect\citeauthoryear{Jeon, Kim, Lee, Kang, and Lee}{Jeon
  et~al\mbox{.}}{2022}]%
        {JeonKLKL22}
\bibfield{author}{\bibinfo{person}{Myeongho Jeon}, \bibinfo{person}{Daekyung
  Kim}, \bibinfo{person}{Woochul Lee}, \bibinfo{person}{Myungjoo Kang}, {and}
  \bibinfo{person}{Joonseok Lee}.} \bibinfo{year}{2022}\natexlab{}.
\newblock \showarticletitle{A Conservative Approach for Unbiased Learning on
  Unknown Biases}. In \bibinfo{booktitle}{\emph{CVPR}}.
\newblock


\bibitem[\protect\citeauthoryear{Joachims, Swaminathan, and Schnabel}{Joachims
  et~al\mbox{.}}{2017}]%
        {JoachimsSS17}
\bibfield{author}{\bibinfo{person}{Thorsten Joachims}, \bibinfo{person}{Adith
  Swaminathan}, {and} \bibinfo{person}{Tobias Schnabel}.}
  \bibinfo{year}{2017}\natexlab{}.
\newblock \showarticletitle{Unbiased Learning-to-Rank with Biased Feedback}. In
  \bibinfo{booktitle}{\emph{WSDM}}. \bibinfo{pages}{781--789}.
\newblock


\bibitem[\protect\citeauthoryear{Kingma and Ba}{Kingma and Ba}{2015}]%
        {KingmaB14}
\bibfield{author}{\bibinfo{person}{Diederik~P. Kingma} {and}
  \bibinfo{person}{Jimmy Ba}.} \bibinfo{year}{2015}\natexlab{}.
\newblock \showarticletitle{Adam: {A} Method for Stochastic Optimization}. In
  \bibinfo{booktitle}{\emph{ICLR}}.
\newblock


\bibitem[\protect\citeauthoryear{Lee, Kim, Lebanon, and Singer}{Lee
  et~al\mbox{.}}{2013}]%
        {LeeKLS13}
\bibfield{author}{\bibinfo{person}{Joonseok Lee}, \bibinfo{person}{Seungyeon
  Kim}, \bibinfo{person}{Guy Lebanon}, {and} \bibinfo{person}{Yoram Singer}.}
  \bibinfo{year}{2013}\natexlab{}.
\newblock \showarticletitle{Local low-rank matrix approximation}. In
  \bibinfo{booktitle}{\emph{ICML}}. \bibinfo{pages}{82--90}.
\newblock


\bibitem[\protect\citeauthoryear{Lee, Park, and Lee}{Lee et~al\mbox{.}}{2021}]%
        {LeePL21}
\bibfield{author}{\bibinfo{person}{Jae{-}woong Lee}, \bibinfo{person}{Seongmin
  Park}, {and} \bibinfo{person}{Jongwuk Lee}.} \bibinfo{year}{2021}\natexlab{}.
\newblock \showarticletitle{Dual Unbiased Recommender Learning for Implicit
  Feedback}. In \bibinfo{booktitle}{\emph{{SIGIR}}}.
  \bibinfo{pages}{1647--1651}.
\newblock


\bibitem[\protect\citeauthoryear{Lee, Song, Haam, Lee, Choi, and Lee}{Lee
  et~al\mbox{.}}{2020}]%
        {LeeSHLCL20}
\bibfield{author}{\bibinfo{person}{Jae{-}woong Lee},
  \bibinfo{person}{Young{-}In Song}, \bibinfo{person}{Deokmin Haam},
  \bibinfo{person}{Sanghoon Lee}, \bibinfo{person}{Woo{-}Sik Choi}, {and}
  \bibinfo{person}{Jongwuk Lee}.} \bibinfo{year}{2020}\natexlab{}.
\newblock \showarticletitle{Bridging the Gap between Click and Relevance for
  Learning-to-Rank with Minimal Supervision}. In
  \bibinfo{booktitle}{\emph{{CIKM}}}. \bibinfo{publisher}{{ACM}},
  \bibinfo{pages}{2109--2112}.
\newblock


\bibitem[\protect\citeauthoryear{Liang, Charlin, McInerney, and Blei}{Liang
  et~al\mbox{.}}{2016}]%
        {LiangCMB16}
\bibfield{author}{\bibinfo{person}{Dawen Liang}, \bibinfo{person}{Laurent
  Charlin}, \bibinfo{person}{James McInerney}, {and} \bibinfo{person}{David~M.
  Blei}.} \bibinfo{year}{2016}\natexlab{}.
\newblock \showarticletitle{Modeling User Exposure in Recommendation}. In
  \bibinfo{booktitle}{\emph{WWW}}. \bibinfo{pages}{951--961}.
\newblock


\bibitem[\protect\citeauthoryear{Lipton, Wang, and Smola}{Lipton
  et~al\mbox{.}}{2018}]%
        {LiptonWS18}
\bibfield{author}{\bibinfo{person}{Zachary~C. Lipton},
  \bibinfo{person}{Yu{-}Xiang Wang}, {and} \bibinfo{person}{Alexander~J.
  Smola}.} \bibinfo{year}{2018}\natexlab{}.
\newblock \showarticletitle{Detecting and Correcting for Label Shift with Black
  Box Predictors}. In \bibinfo{booktitle}{\emph{{ICML}}}
  \emph{(\bibinfo{series}{Proceedings of Machine Learning Research},
  Vol.~\bibinfo{volume}{80})}. \bibinfo{publisher}{{PMLR}},
  \bibinfo{pages}{3128--3136}.
\newblock


\bibitem[\protect\citeauthoryear{Ma and Chen}{Ma and Chen}{2019}]%
        {MaC19}
\bibfield{author}{\bibinfo{person}{Wei Ma} {and} \bibinfo{person}{George~H.
  Chen}.} \bibinfo{year}{2019}\natexlab{}.
\newblock \showarticletitle{Missing Not at Random in Matrix Completion: The
  Effectiveness of Estimating Missingness Probabilities Under a Low Nuclear
  Norm Assumption}. In \bibinfo{booktitle}{\emph{NeurIPS}}.
  \bibinfo{pages}{14871--14880}.
\newblock


\bibitem[\protect\citeauthoryear{Marlin and Zemel}{Marlin and Zemel}{2009}]%
        {MarlinZ09}
\bibfield{author}{\bibinfo{person}{Benjamin~M. Marlin} {and}
  \bibinfo{person}{Richard~S. Zemel}.} \bibinfo{year}{2009}\natexlab{}.
\newblock \showarticletitle{Collaborative prediction and ranking with
  non-random missing data}. In \bibinfo{booktitle}{\emph{RecSys}}.
  \bibinfo{pages}{5--12}.
\newblock


\bibitem[\protect\citeauthoryear{Marlin, Zemel, Roweis, and Slaney}{Marlin
  et~al\mbox{.}}{2007}]%
        {MarlinZRS07}
\bibfield{author}{\bibinfo{person}{Benjamin~M. Marlin},
  \bibinfo{person}{Richard~S. Zemel}, \bibinfo{person}{Sam~T. Roweis}, {and}
  \bibinfo{person}{Malcolm Slaney}.} \bibinfo{year}{2007}\natexlab{}.
\newblock \showarticletitle{Collaborative Filtering and the Missing at Random
  Assumption}. In \bibinfo{booktitle}{\emph{UAI}}. \bibinfo{pages}{267--275}.
\newblock


\bibitem[\protect\citeauthoryear{Mobahi, Farajtabar, and Bartlett}{Mobahi
  et~al\mbox{.}}{2020}]%
        {MobahiFB20}
\bibfield{author}{\bibinfo{person}{Hossein Mobahi}, \bibinfo{person}{Mehrdad
  Farajtabar}, {and} \bibinfo{person}{Peter~L. Bartlett}.}
  \bibinfo{year}{2020}\natexlab{}.
\newblock \showarticletitle{Self-Distillation Amplifies Regularization in
  Hilbert Space}. In \bibinfo{booktitle}{\emph{NeurIPS}}.
\newblock


\bibitem[\protect\citeauthoryear{Ovaisi, Ahsan, Zhang, Vasilaky, and
  Zheleva}{Ovaisi et~al\mbox{.}}{2020}]%
        {OvaisiAZVZ20}
\bibfield{author}{\bibinfo{person}{Zohreh Ovaisi}, \bibinfo{person}{Ragib
  Ahsan}, \bibinfo{person}{Yifan Zhang}, \bibinfo{person}{Kathryn Vasilaky},
  {and} \bibinfo{person}{Elena Zheleva}.} \bibinfo{year}{2020}\natexlab{}.
\newblock \showarticletitle{Correcting for Selection Bias in Learning-to-rank
  Systems}. In \bibinfo{booktitle}{\emph{{WWW}}}. \bibinfo{pages}{1863--1873}.
\newblock


\bibitem[\protect\citeauthoryear{Prathama, Senjaya, Yahya, and Wu}{Prathama
  et~al\mbox{.}}{2021}]%
        {PrathamaSYW21}
\bibfield{author}{\bibinfo{person}{Frans Prathama},
  \bibinfo{person}{Wenny~Franciska Senjaya}, \bibinfo{person}{Bernardo~Nugroho
  Yahya}, {and} \bibinfo{person}{Jei{-}Zheng Wu}.}
  \bibinfo{year}{2021}\natexlab{}.
\newblock \showarticletitle{Personalized recommendation by matrix
  co-factorization with multiple implicit feedback on pairwise comparison}.
\newblock \bibinfo{journal}{\emph{Comput. Ind. Eng.}}  \bibinfo{volume}{152}
  (\bibinfo{year}{2021}), \bibinfo{pages}{107033}.
\newblock


\bibitem[\protect\citeauthoryear{Qin, Chen, Metzler, Noh, Qin, and Wang}{Qin
  et~al\mbox{.}}{2020}]%
        {QinCMNQW20}
\bibfield{author}{\bibinfo{person}{Zhen Qin}, \bibinfo{person}{Suming~J. Chen},
  \bibinfo{person}{Donald Metzler}, \bibinfo{person}{Yongwoo Noh},
  \bibinfo{person}{Jingzheng Qin}, {and} \bibinfo{person}{Xuanhui Wang}.}
  \bibinfo{year}{2020}\natexlab{}.
\newblock \showarticletitle{Attribute-Based Propensity for Unbiased Learning in
  Recommender Systems: Algorithm and Case Studies}. In
  \bibinfo{booktitle}{\emph{KDD}}. \bibinfo{pages}{2359–2367}.
\newblock


\bibitem[\protect\citeauthoryear{Rendle}{Rendle}{2021}]%
        {Rendle21}
\bibfield{author}{\bibinfo{person}{Steffen Rendle}.}
  \bibinfo{year}{2021}\natexlab{}.
\newblock \showarticletitle{Item Recommendation from Implicit Feedback}.
\newblock \bibinfo{journal}{\emph{arXiv:2101.08769}} (\bibinfo{year}{2021}).
\newblock


\bibitem[\protect\citeauthoryear{Rendle, Freudenthaler, Gantner, and
  Schmidt{-}Thieme}{Rendle et~al\mbox{.}}{2009}]%
        {RendleFGS09}
\bibfield{author}{\bibinfo{person}{Steffen Rendle}, \bibinfo{person}{Christoph
  Freudenthaler}, \bibinfo{person}{Zeno Gantner}, {and} \bibinfo{person}{Lars
  Schmidt{-}Thieme}.} \bibinfo{year}{2009}\natexlab{}.
\newblock \showarticletitle{{BPR:} Bayesian Personalized Ranking from Implicit
  Feedback}. In \bibinfo{booktitle}{\emph{UAI}}. \bibinfo{pages}{452--461}.
\newblock


\bibitem[\protect\citeauthoryear{Ricci, Rokach, and Shapira}{Ricci
  et~al\mbox{.}}{2015}]%
        {RicciRS15}
\bibfield{editor}{\bibinfo{person}{Francesco Ricci}, \bibinfo{person}{Lior
  Rokach}, {and} \bibinfo{person}{Bracha Shapira}} (Eds.).
  \bibinfo{year}{2015}\natexlab{}.
\newblock \bibinfo{booktitle}{\emph{Recommender Systems Handbook}}.
\newblock \bibinfo{publisher}{Springer}.
\newblock
\showISBNx{978-0-387-85820-3}


\bibitem[\protect\citeauthoryear{Saito}{Saito}{2020a}]%
        {Saito20at}
\bibfield{author}{\bibinfo{person}{Yuta Saito}.}
  \bibinfo{year}{2020}\natexlab{a}.
\newblock \showarticletitle{Asymmetric Tri-training for Debiasing
  Missing-Not-At-Random Explicit Feedback}. In
  \bibinfo{booktitle}{\emph{SIGIR}}. \bibinfo{pages}{309--318}.
\newblock


\bibitem[\protect\citeauthoryear{Saito}{Saito}{2020b}]%
        {Saito20}
\bibfield{author}{\bibinfo{person}{Yuta Saito}.}
  \bibinfo{year}{2020}\natexlab{b}.
\newblock \showarticletitle{Unbiased Pairwise Learning from Biased Implicit
  Feedback}. In \bibinfo{booktitle}{\emph{ICTIR}}. \bibinfo{pages}{5--12}.
\newblock


\bibitem[\protect\citeauthoryear{Saito, Yaginuma, Nishino, Sakata, and
  Nakata}{Saito et~al\mbox{.}}{2020}]%
        {SaitoYNSN20}
\bibfield{author}{\bibinfo{person}{Yuta Saito}, \bibinfo{person}{Suguru
  Yaginuma}, \bibinfo{person}{Yuta Nishino}, \bibinfo{person}{Hayato Sakata},
  {and} \bibinfo{person}{Kazuhide Nakata}.} \bibinfo{year}{2020}\natexlab{}.
\newblock \showarticletitle{Unbiased Recommender Learning from
  Missing-Not-At-Random Implicit Feedback}. In
  \bibinfo{booktitle}{\emph{WSDM}}. \bibinfo{pages}{501--509}.
\newblock


\bibitem[\protect\citeauthoryear{Schnabel, Swaminathan, Singh, Chandak, and
  Joachims}{Schnabel et~al\mbox{.}}{2016}]%
        {SchnabelSSCJ16}
\bibfield{author}{\bibinfo{person}{Tobias Schnabel}, \bibinfo{person}{Adith
  Swaminathan}, \bibinfo{person}{Ashudeep Singh}, \bibinfo{person}{Navin
  Chandak}, {and} \bibinfo{person}{Thorsten Joachims}.}
  \bibinfo{year}{2016}\natexlab{}.
\newblock \showarticletitle{Recommendations as Treatments: Debiasing Learning
  and Evaluation}. In \bibinfo{booktitle}{\emph{ICML}}.
  \bibinfo{pages}{1670--1679}.
\newblock


\bibitem[\protect\citeauthoryear{Sedhain, Menon, Sanner, and Xie}{Sedhain
  et~al\mbox{.}}{2015}]%
        {SedhainMSX15}
\bibfield{author}{\bibinfo{person}{Suvash Sedhain},
  \bibinfo{person}{Aditya~Krishna Menon}, \bibinfo{person}{Scott Sanner}, {and}
  \bibinfo{person}{Lexing Xie}.} \bibinfo{year}{2015}\natexlab{}.
\newblock \showarticletitle{Autorec: Autoencoders meet collaborative
  filtering}. In \bibinfo{booktitle}{\emph{WWW}}. \bibinfo{pages}{111--112}.
\newblock


\bibitem[\protect\citeauthoryear{Shenbin, Alekseev, Tutubalina, Malykh, and
  Nikolenko}{Shenbin et~al\mbox{.}}{2020}]%
        {ShenbinATMN20}
\bibfield{author}{\bibinfo{person}{Ilya Shenbin}, \bibinfo{person}{Anton
  Alekseev}, \bibinfo{person}{Elena Tutubalina}, \bibinfo{person}{Valentin
  Malykh}, {and} \bibinfo{person}{Sergey~I. Nikolenko}.}
  \bibinfo{year}{2020}\natexlab{}.
\newblock \showarticletitle{RecVAE: {A} New Variational Autoencoder for Top-N
  Recommendations with Implicit Feedback}. In
  \bibinfo{booktitle}{\emph{{WSDM}}}. \bibinfo{pages}{528--536}.
\newblock


\bibitem[\protect\citeauthoryear{Steck}{Steck}{2010}]%
        {Steck10}
\bibfield{author}{\bibinfo{person}{Harald Steck}.}
  \bibinfo{year}{2010}\natexlab{}.
\newblock \showarticletitle{Training and testing of recommender systems on data
  missing not at random}. In \bibinfo{booktitle}{\emph{KDD}}.
  \bibinfo{pages}{713--722}.
\newblock


\bibitem[\protect\citeauthoryear{Swaminathan and Joachims}{Swaminathan and
  Joachims}{2015}]%
        {SwaminathanJ15}
\bibfield{author}{\bibinfo{person}{Adith Swaminathan} {and}
  \bibinfo{person}{Thorsten Joachims}.} \bibinfo{year}{2015}\natexlab{}.
\newblock \showarticletitle{The Self-Normalized Estimator for Counterfactual
  Learning}. In \bibinfo{booktitle}{\emph{NIPS}}. \bibinfo{pages}{3231–3239}.
\newblock


\bibitem[\protect\citeauthoryear{Tang and Wang}{Tang and Wang}{2018}]%
        {TangW18a}
\bibfield{author}{\bibinfo{person}{Jiaxi Tang} {and} \bibinfo{person}{Ke
  Wang}.} \bibinfo{year}{2018}\natexlab{}.
\newblock \showarticletitle{Personalized Top-N Sequential Recommendation via
  Convolutional Sequence Embedding}. In \bibinfo{booktitle}{\emph{WSDM}}.
  \bibinfo{pages}{565--573}.
\newblock


\bibitem[\protect\citeauthoryear{Wang, Wang, and Yeung}{Wang
  et~al\mbox{.}}{2015}]%
        {WangWY15}
\bibfield{author}{\bibinfo{person}{Hao Wang}, \bibinfo{person}{Naiyan Wang},
  {and} \bibinfo{person}{Dit{-}Yan Yeung}.} \bibinfo{year}{2015}\natexlab{}.
\newblock \showarticletitle{Collaborative Deep Learning for Recommender
  Systems}. In \bibinfo{booktitle}{\emph{KDD}}. \bibinfo{pages}{1235--1244}.
\newblock


\bibitem[\protect\citeauthoryear{Wang, de~Vries, and Reinders}{Wang
  et~al\mbox{.}}{2006}]%
        {WangVR06}
\bibfield{author}{\bibinfo{person}{Jun Wang}, \bibinfo{person}{Arjen~P. de
  Vries}, {and} \bibinfo{person}{Marcel J.~T. Reinders}.}
  \bibinfo{year}{2006}\natexlab{}.
\newblock \showarticletitle{Unifying user-based and item-based collaborative
  filtering approaches by similarity fusion}. In
  \bibinfo{booktitle}{\emph{{SIGIR}}}. \bibinfo{pages}{501--508}.
\newblock


\bibitem[\protect\citeauthoryear{Wang, Qin, Wang, and Wang}{Wang
  et~al\mbox{.}}{2021a}]%
        {Wang0WW21}
\bibfield{author}{\bibinfo{person}{Nan Wang}, \bibinfo{person}{Zhen Qin},
  \bibinfo{person}{Xuanhui Wang}, {and} \bibinfo{person}{Hongning Wang}.}
  \bibinfo{year}{2021}\natexlab{a}.
\newblock \showarticletitle{Non-Clicks Mean Irrelevant? Propensity Ratio
  Scoring As a Correction}. In \bibinfo{booktitle}{\emph{{WSDM}}}.
  \bibinfo{pages}{481--489}.
\newblock


\bibitem[\protect\citeauthoryear{Wang, Bendersky, Metzler, and Najork}{Wang
  et~al\mbox{.}}{2016}]%
        {WangBMN16}
\bibfield{author}{\bibinfo{person}{Xuanhui Wang}, \bibinfo{person}{Michael
  Bendersky}, \bibinfo{person}{Donald Metzler}, {and} \bibinfo{person}{Marc
  Najork}.} \bibinfo{year}{2016}\natexlab{}.
\newblock \showarticletitle{Learning to Rank with Selection Bias in Personal
  Search}. In \bibinfo{booktitle}{\emph{SIGIR}}. \bibinfo{pages}{115--124}.
\newblock


\bibitem[\protect\citeauthoryear{Wang, Golbandi, Bendersky, Metzler, and
  Najork}{Wang et~al\mbox{.}}{2018a}]%
        {WangGBMN18}
\bibfield{author}{\bibinfo{person}{Xuanhui Wang}, \bibinfo{person}{Nadav
  Golbandi}, \bibinfo{person}{Michael Bendersky}, \bibinfo{person}{Donald
  Metzler}, {and} \bibinfo{person}{Marc Najork}.}
  \bibinfo{year}{2018}\natexlab{a}.
\newblock \showarticletitle{Position Bias Estimation for Unbiased Learning to
  Rank in Personal Search}. In \bibinfo{booktitle}{\emph{WSDM}}.
  \bibinfo{pages}{610--618}.
\newblock


\bibitem[\protect\citeauthoryear{Wang, Zhang, Sun, and Qi}{Wang
  et~al\mbox{.}}{2019}]%
        {WangZSQ19}
\bibfield{author}{\bibinfo{person}{Xiaojie Wang}, \bibinfo{person}{Rui Zhang},
  \bibinfo{person}{Yu Sun}, {and} \bibinfo{person}{Jianzhong Qi}.}
  \bibinfo{year}{2019}\natexlab{}.
\newblock \showarticletitle{Doubly Robust Joint Learning for Recommendation on
  Data Missing Not at Random}. In \bibinfo{booktitle}{\emph{ICML}}.
  \bibinfo{pages}{6638--6647}.
\newblock


\bibitem[\protect\citeauthoryear{Wang, Zhang, Sun, and Qi}{Wang
  et~al\mbox{.}}{2021b}]%
        {0003ZS021}
\bibfield{author}{\bibinfo{person}{Xiaojie Wang}, \bibinfo{person}{Rui Zhang},
  \bibinfo{person}{Yu Sun}, {and} \bibinfo{person}{Jianzhong Qi}.}
  \bibinfo{year}{2021}\natexlab{b}.
\newblock \showarticletitle{Combating Selection Biases in Recommender Systems
  with a Few Unbiased Ratings}. In \bibinfo{booktitle}{\emph{{WSDM}}}.
  \bibinfo{pages}{427--435}.
\newblock


\bibitem[\protect\citeauthoryear{Wang, Liang, Charlin, and Blei}{Wang
  et~al\mbox{.}}{2018b}]%
        {abs-1808-06581}
\bibfield{author}{\bibinfo{person}{Yixin Wang}, \bibinfo{person}{Dawen Liang},
  \bibinfo{person}{Laurent Charlin}, {and} \bibinfo{person}{David~M. Blei}.}
  \bibinfo{year}{2018}\natexlab{b}.
\newblock \showarticletitle{The Deconfounded Recommender: {A} Causal Inference
  Approach to Recommendation}.
\newblock \bibinfo{journal}{\emph{CoRR}}  \bibinfo{volume}{abs/1808.06581}
  (\bibinfo{year}{2018}).
\newblock


\bibitem[\protect\citeauthoryear{Wei, Feng, Chen, Wu, Yi, and He}{Wei
  et~al\mbox{.}}{2021}]%
        {WeiFCWYH21}
\bibfield{author}{\bibinfo{person}{Tianxin Wei}, \bibinfo{person}{Fuli Feng},
  \bibinfo{person}{Jiawei Chen}, \bibinfo{person}{Ziwei Wu},
  \bibinfo{person}{Jinfeng Yi}, {and} \bibinfo{person}{Xiangnan He}.}
  \bibinfo{year}{2021}\natexlab{}.
\newblock \showarticletitle{Model-agnostic counterfactual reasoning for
  eliminating popularity bias in recommender system}. In
  \bibinfo{booktitle}{\emph{KDD}}. \bibinfo{pages}{1791--1800}.
\newblock


\bibitem[\protect\citeauthoryear{Wu, DuBois, Zheng, and Ester}{Wu
  et~al\mbox{.}}{2016}]%
        {WuDZE16}
\bibfield{author}{\bibinfo{person}{Yao Wu}, \bibinfo{person}{Christopher
  DuBois}, \bibinfo{person}{Alice~X. Zheng}, {and} \bibinfo{person}{Martin
  Ester}.} \bibinfo{year}{2016}\natexlab{}.
\newblock \showarticletitle{Collaborative Denoising Auto-Encoders for Top-N
  Recommender Systems}. In \bibinfo{booktitle}{\emph{{WSDM}}}.
  \bibinfo{pages}{153--162}.
\newblock


\bibitem[\protect\citeauthoryear{Xue, Dai, Zhang, Huang, and Chen}{Xue
  et~al\mbox{.}}{2017}]%
        {XueDZHC17}
\bibfield{author}{\bibinfo{person}{Hong{-}Jian Xue}, \bibinfo{person}{Xinyu
  Dai}, \bibinfo{person}{Jianbing Zhang}, \bibinfo{person}{Shujian Huang},
  {and} \bibinfo{person}{Jiajun Chen}.} \bibinfo{year}{2017}\natexlab{}.
\newblock \showarticletitle{Deep Matrix Factorization Models for Recommender
  Systems}. In \bibinfo{booktitle}{\emph{IJCAI}}. \bibinfo{pages}{3203--3209}.
\newblock


\bibitem[\protect\citeauthoryear{Yamashita, Kawamura, and Suzuki}{Yamashita
  et~al\mbox{.}}{2011}]%
        {YamashitaKS11}
\bibfield{author}{\bibinfo{person}{Akihiro Yamashita},
  \bibinfo{person}{Hidenori Kawamura}, {and} \bibinfo{person}{Keiji Suzuki}.}
  \bibinfo{year}{2011}\natexlab{}.
\newblock \showarticletitle{Adaptive Fusion Method for User-Based and
  Item-Based Collaborative Filtering}.
\newblock \bibinfo{journal}{\emph{Adv. Complex Syst.}} \bibinfo{volume}{14},
  \bibinfo{number}{2} (\bibinfo{year}{2011}), \bibinfo{pages}{133--149}.
\newblock


\bibitem[\protect\citeauthoryear{Yang, Xie, Su, and Yuille}{Yang
  et~al\mbox{.}}{2019}]%
        {YangXSY19}
\bibfield{author}{\bibinfo{person}{Chenglin Yang}, \bibinfo{person}{Lingxi
  Xie}, \bibinfo{person}{Chi Su}, {and} \bibinfo{person}{Alan~L. Yuille}.}
  \bibinfo{year}{2019}\natexlab{}.
\newblock \showarticletitle{Snapshot Distillation: Teacher-Student Optimization
  in One Generation}. In \bibinfo{booktitle}{\emph{{CVPR}}}.
  \bibinfo{publisher}{Computer Vision Foundation / {IEEE}},
  \bibinfo{pages}{2859--2868}.
\newblock


\bibitem[\protect\citeauthoryear{Yang, Cui, Xuan, Wang, Belongie, and
  Estrin}{Yang et~al\mbox{.}}{2018}]%
        {YangCXWBE18}
\bibfield{author}{\bibinfo{person}{Longqi Yang}, \bibinfo{person}{Yin Cui},
  \bibinfo{person}{Yuan Xuan}, \bibinfo{person}{Chenyang Wang},
  \bibinfo{person}{Serge Belongie}, {and} \bibinfo{person}{Deborah Estrin}.}
  \bibinfo{year}{2018}\natexlab{}.
\newblock \showarticletitle{Unbiased Offline Recommender Evaluation for
  Missing-Not-at-Random Implicit Feedback}. In
  \bibinfo{booktitle}{\emph{RecSys}}. \bibinfo{pages}{279–287}.
\newblock


\bibitem[\protect\citeauthoryear{Yao, Tong, Yan, Xu, Zhang, Szymanski, and
  Lu}{Yao et~al\mbox{.}}{2019}]%
        {YaoTYXZSL19}
\bibfield{author}{\bibinfo{person}{Yuan Yao}, \bibinfo{person}{Hanghang Tong},
  \bibinfo{person}{Guo Yan}, \bibinfo{person}{Feng Xu}, \bibinfo{person}{Xiang
  Zhang}, \bibinfo{person}{Boleslaw~K. Szymanski}, {and} \bibinfo{person}{Jian
  Lu}.} \bibinfo{year}{2019}\natexlab{}.
\newblock \showarticletitle{Dual-regularized one-class collaborative filtering
  with implicit feedback}.
\newblock \bibinfo{journal}{\emph{World Wide Web}} \bibinfo{volume}{22},
  \bibinfo{number}{3} (\bibinfo{year}{2019}), \bibinfo{pages}{1099--1129}.
\newblock


\bibitem[\protect\citeauthoryear{Zhang, Feng, He, Wei, Song, Ling, and
  Zhang}{Zhang et~al\mbox{.}}{2021}]%
        {ZhangFHWSLZ21}
\bibfield{author}{\bibinfo{person}{Yang Zhang}, \bibinfo{person}{Fuli Feng},
  \bibinfo{person}{Xiangnan He}, \bibinfo{person}{Tianxin Wei},
  \bibinfo{person}{Chonggang Song}, \bibinfo{person}{Guohui Ling}, {and}
  \bibinfo{person}{Yongdong Zhang}.} \bibinfo{year}{2021}\natexlab{}.
\newblock \showarticletitle{Causal Intervention for Leveraging Popularity Bias
  in Recommendation}.
\newblock \bibinfo{journal}{\emph{SIGIR}} (\bibinfo{year}{2021}),
  \bibinfo{pages}{11--20}.
\newblock


\bibitem[\protect\citeauthoryear{Zheng, Gao, Li, He, Li, and Jin}{Zheng
  et~al\mbox{.}}{2021}]%
        {ZhengGLHLJ21}
\bibfield{author}{\bibinfo{person}{Yu Zheng}, \bibinfo{person}{Chen Gao},
  \bibinfo{person}{Xiang Li}, \bibinfo{person}{Xiangnan He},
  \bibinfo{person}{Yong Li}, {and} \bibinfo{person}{Depeng Jin}.}
  \bibinfo{year}{2021}\natexlab{}.
\newblock \showarticletitle{Disentangling User Interest and Conformity for
  Recommendation with Causal Embedding}. In \bibinfo{booktitle}{\emph{WWW}}.
  \bibinfo{pages}{2980--2991}.
\newblock


\bibitem[\protect\citeauthoryear{Zhu, He, Zhang, and Caverlee}{Zhu
  et~al\mbox{.}}{2020}]%
        {ZhuHZC20}
\bibfield{author}{\bibinfo{person}{Ziwei Zhu}, \bibinfo{person}{Yun He},
  \bibinfo{person}{Yin Zhang}, {and} \bibinfo{person}{James Caverlee}.}
  \bibinfo{year}{2020}\natexlab{}.
\newblock \showarticletitle{Unbiased Implicit Recommendation and Propensity
  Estimation via Combinational Joint Learning}. In
  \bibinfo{booktitle}{\emph{RecSys}}. \bibinfo{pages}{551–556}.
\newblock


\bibitem[\protect\citeauthoryear{Zhu, Wang, and Caverlee}{Zhu
  et~al\mbox{.}}{2019}]%
        {ZhuWC19}
\bibfield{author}{\bibinfo{person}{Ziwei Zhu}, \bibinfo{person}{Jianling Wang},
  {and} \bibinfo{person}{James Caverlee}.} \bibinfo{year}{2019}\natexlab{}.
\newblock \showarticletitle{Improving Top-K Recommendation via
  JointCollaborative Autoencoders}. In \bibinfo{booktitle}{\emph{{WWW}}}.
  \bibinfo{pages}{3483--3482}.
\newblock


\end{thebibliography}

\end{document}